\documentclass[
aps, prd,  notitlepage, 
floats, floatfix, onecolumn,
amsmath, amssymb, amsfonts, eqsecnum,
superscriptaddress,
showpacs, showkeys,
nofootinbib,
longbibliography,
]{revtex4-2}
\usepackage{multirow}
\usepackage{graphicx}
\usepackage{subfigure}
\usepackage[usenames,dvipsnames]{xcolor}
\usepackage{tikz}
\usepackage{xspace} 
\usepackage{bm}
\usepackage[utf8]{inputenc} 
\usepackage{latexsym}
\usepackage{float}
\usepackage{ulem}

\usepackage[utf8]{inputenc} 
\DeclareUnicodeCharacter{200B}{\hskip 0pt}

\usepackage{diagbox} 
\usepackage{makecell}
\usepackage{multirow}

\DeclareMathOperator*{\argmax}{argmax}

\renewcommand{\emph}[1]{\textit{#1}}
\xdefinecolor{mylinkcolor}{rgb}{0,0,0.5}
\usepackage[
bookmarksnumbered, bookmarksopen, bookmarksopenlevel=2,
breaklinks=true,
colorlinks=true, filecolor=mylinkcolor, citecolor=mylinkcolor,
linkcolor=mylinkcolor, urlcolor=mylinkcolor, menucolor=mylinkcolor,
]{hyperref}

\def\be{\begin{equation}}
	\def\ee{\end{equation}}
\def\bea{\begin{eqnarray}}
	\def\eea{\end{eqnarray}}
\newcommand{\bes}{\begin{subequations}}
	\newcommand{\ees}{\end{subequations}}
\newcommand{\besa}{\begin{subequations} \begin{align}}
		\newcommand{\eesa}{\end{align} \end{subequations}}
\def\comment#1{}

\usepackage{color}

\allowdisplaybreaks

\begin{document}
	\bibliographystyle{unsrt}
	\title{Constraining the Deviation of Kerr Metric via Bumpy Parameterization and Particle Swarm Optimization in Extreme Mass-Ratio Inspirals}
	
	\author{Xiaobo Zou}
	\affiliation{School of Fundamental Physics and Mathematical Sciences, Hangzhou Institute for Advanced Study, UCAS, Hangzhou 310024, China}
	
	\author{Xingyu Zhong}
	\affiliation{School of Fundamental Physics and Mathematical Sciences, Hangzhou Institute for Advanced Study, UCAS, Hangzhou 310024, China}

	\author{Wen-Biao Han}
	\email{Conrresponding author: wbhan@shao.ac.cn}
	\affiliation{Shanghai Astronomical Observatory,  Chinese Academy of Sciences,  Shanghai,  China,  200030}
	\affiliation{School of Fundamental Physics and Mathematical Sciences, Hangzhou Institute for Advanced Study, UCAS, Hangzhou 310024, China}
	\affiliation{Taiji Laboratory for Gravitational Wave Universe (Beijing/Hangzhou), University of Chinese Academy of Sciences, Beijing 100049, China}
	\affiliation{School of Astronomy and Space Science, University of Chinese Academy of Sciences, Beijing 100049, China}
	\affiliation{State Key Laboratory of Radio Astronomy and Technology, A20 Datun Road, Chaoyang District, Beijing, 100101, P. R. China}
	\author{Soumya D. Mohanty}
	\email{Conrresponding author: soumya.mohanty@utrgv.edu}
	\affiliation{Department of Physics and Astronomy, The University of Texas Rio Grande Valley,Brownsville, TX 78520, USA}
	
	
\begin{abstract}
Measurement of deviations in the Kerr metric using gravitational wave (GW) observations will provide a clear signal of new Physics. Previous studies have developed multiple parameterizations (e.g. ``bumpy" spacetime) to characterize such deviations in extreme mass ratio inspirals (EMRI). These approaches often rely on the Fisher information matrix (FIM) formalism to quantify the constraining power of future space-borne gravitational-wave detectors, such as LISA and Tianqin. For instance, using the analytical kludge waveform model under varying source configurations, such methods have achieved constraint sensitivity levels ranging from $10^{-4} \sim 10^{-2}$ on the dimensionless bumpy parameter $\delta \tilde{Q}$ for LISA. In this paper, we advance prior analyses by integrating particle swarm optimization (PSO) with matched filtering under a restricted parameter search range to enforce a high probability of convergence for PSO. Our results reveal a significant number of degenerate peaks in the likelihood function over the signal parameter space with values that exceed the injected one. 
This extreme level of degeneracy arises from the involvement of the additional bumpy parameter $\delta \tilde{Q}$ in the parameter space and introduces systematic errors in parameter estimation. We show that these systematic errors can be mitigated using information contained in the ensemble of degenerate peaks, thereby showing a promising potential method for improving local parameter estimation if the other immense challenges of a global search are first solved.
This study highlights the critical importance of accounting for such degeneracies, which are absent in FIM-based analyses, and points out future directions for improving EMRI data analysis.
\end{abstract}
\maketitle
	
\section{Introduction}\label{intro}
Extreme mass-ratio inspirals (EMRI) represent key observational targets for space-borne gravitational wave (GW) detectors such as LISA~\cite{LISA:2017pwj}, Taiji~\cite{Hu:2017mde}, and Tianqin~\cite{TianQin:2015yph}. These systems involve a compact object (CO) orbiting a massive black hole (MBH) with mass ratios in the range $[10^4, 10^7]$, emitting GWs during their inspiral phase~\cite{Fan:2020zhy,Babak:2017tow}. The millihertz-frequency GWs from a year-long observation prior to the final plunge contain over $10^5$ waveform cycles~\cite{Gair:2004iv}, encoding rich phase evolution information. This enables precise parameter estimation of MBH masses and spins to an accuracy of $10^{-5}$ for LISA~\cite{Barack:2003fp}, making EMRIs ideal laboratories for probing strong-field gravity~\cite{Glampedakis:2005cf}, testing general relativity (GR)~\cite{Chua:2018yng,Shen:2023pje,Kumar:2024utz}, constraining MBH/CO properties~\cite{Zi:2021pdp,Barsanti:2022vvl,Huerta:2011kt,Cui:2025bgu}, and studying galactic environmental effects~\cite{Destounis:2022obl}.
	
Building on foundational works by Ryan~\cite{Ryan:1995wh,Ryan:1997hg}, which mapped the multipole moments of kerr MBH to GW phases for equatorial and circular EMRIs, subsequent studies generalized this framework to inclined and eccentric orbits~\cite{Barack:2006pq}. Further extensions incorporated MBH motion, higher-order multipole moments, and tidal effects~\cite{Datta:2019euh}, establishing the theoretical basis for constraining intrinsic MBH properties via EMRI GWs. Of particular interest are parameterized deviations from the kerr metric~\cite{AbhishekChowdhuri:2023gvu}, a key prediction of GR in the strong-field regime. Several parameterizations~\cite{Glampedakis:2005cf,Xin:2018urr,AbhishekChowdhuri:2023gvu} have been proposed, including the ``bumpy" Kerr formalism~\cite{Barack:2006pq,Collins:2004ex,Vigeland:2011ji,Moore:2017lxy}, which introduces a single parameter $\delta \tilde{Q}$ to quantify quadrupole moment deviations from GR and provide a practical choice for its data analysis.
	
Fisher information matrix (FIM) studies estimate constraints on $\delta \tilde{Q}$ ranging from $[10^{-4}, 10^{-2}]$ for LISA~\cite{Barack:2006pq} to $[10^{-5}, 10^{-1}]$ for Tianqin~\cite{Liu:2020ghq}. However, FIM-predicted errors are asymptotically valid only in the high-SNR regime~\cite{Barack:2003fp,Barack:2006pq,Liu:2020ghq}, contrastingly necessitating empirical simulations incorporating realistic data analysis methods to quantify finite-SNR parameter estimation errors. In this context, matched filtering remains the gold standard for GW data analysis, especially for the moderate to weak SNRs expected for EMRI signals in space-borne detectors.
Early applications of matched filtering to EMRIs emerged in the Mock LISA Data Challenges (MLDC)~\cite{Babak:2008aa,MockLISADataChallengeTaskForce:2009wir}, with recent advances in the LISA Data Challenges (LDC)~\cite{Baghi:2022ucj}. MLDC-era Markov Chain Monte Carlo (MCMC) algorithms employed strategies like top-$3$ harmonics likelihood maximization, tuned proposal distributions, harmonic mapping, and multi-stage search~\cite{Babak:2009ua,Cornish:2008zd,Ali:2012zz}, laying groundwork for EMRI parameter estimation. Subsequent LDC-era efforts introduced either hierarchical semi-coherent searches and  phenomenological waveforms~\cite{Wang:2012xh} to narrow EMRI parameter ranges~\cite{Ye:2023lok,Strub:2025dfs,Cole:2025sqo} or machine learning method for signal detection~\cite{Zhang:2022xuq,Zhao:2023ncy} and parameter estimation~\cite{Yun:2023vwa,Liang:2024qzf,Liang:2025vuf}, the sparse dictionary learning algorithm is also used to speed up the EMRI waveform reconstruction~\cite{Badger:2024rld}. Our prior works~\cite{Zou:2024jqv,Zou:2024osb} demonstrated particle swarm optimization (PSO) as a global optimizer for rapid likelihood maximization, coupled with waveform decomposition and nested optimization to mitigate the high-dimensionality challenges. Recent Taiji data challenges (TDC)~\cite{Du:2025xdq} provide a EMRI dataset including a more realistic orbital dynamics considering the full drag-free and attitude control simulation, and the coupling effect of time-varying armlengths. 
Key challenges in EMRI data analysis include:
\begin{enumerate}
	\item
	\textbf{High-Dimensional Parameter Space}: EMRI waveforms depend on $14$ parameters where $6$ of them  govern the time-dependent phase evolution of GWs, $2$ of them are related with the motion of space-borne detectors while the remaining $6$ modulate the time-independent GW amplitude and phase constants. Their nonlinear coupling complicates simultaneous convergence within computational limits. 
	\item
	\textbf{Needle-in-a-Haystack Waveform Morphology}: Phase coherence renders GWs hypersensitive to phase-coupled parameter variations, creating a sharp $6$-dimensional likelihood peak. While enabling high-precision estimates upon convergence, the broad astrophysical priors(e.g., CO/MBH masses, eccentricity) necessitate costly searches over vast parameter volumes.
	\item
	\textbf{Multimodal Likelihoods}: Waveform harmonics induce secondary likelihood maxima via partial number of harmonics match and parameter degeneracies~\cite{Chua:2021aah,Chua:2022ssg}, obscuring the global maximum corresponding to the true signal.
	\item
	\textbf{Computational Cost}: Generating time-domain waveforms and Time-Delay Interferometry (TDI)~\cite{Tinto:2004wu,LDC-code-maunal} responses for large number of parameter locations remains computationally challenging.
	\end{enumerate}
Due to the fact that longer-duration signals correspond to sharper peak morphology at their location, current LDC studies focus on $0.5\sim1$ year coherent searches~\cite{Ye:2023lok,Strub:2025dfs,Zhang:2022xuq,Zhao:2023ncy,Yun:2023vwa,Badger:2024rld,Zou:2024jqv,Zou:2024osb}, balancing detectability and feasibility. However, extending to LISA’s $2\sim4$ years design lifetime demands hierarchical approaches~\cite{Ye:2023lok} that iteratively narrow parameter ranges, which is a critical area for pipeline development of EMRI data analysis.  Similar challenges are encountered in the search for stellar mass binary black holes in LISA data and hierarchical search techniques developed for such signals~\cite{Bandopadhyay:2023gkb,Fu:2024cpu,Bandopadhyay:2024lwv} could inform those for EMRI signals.

Incorporating the bumpy parameter $\delta \tilde{Q}$ increases the parameter space to $15$ dimensions, exacerbating both search complexity ($7$ phase-coupled parameters corresponding to a $7$-dimensional sharp peak for its waveform morphology) and parameter correlations. Previous Bayesian studies assumed narrow priors to mitigate the needle-in-a-Haystack challenge for a coherent search, testing GR via template mismodeling~\cite{Shen:2025svs} and scalar charge effects~\cite{Speri:2024qak}. In this work, we inherit the setting of narrow search ranges and further incorporate the reduced dimensionality likelihood to alleviate the high-Dimensionality challenge of the full parameter space by extending our AK-focused framework~\cite{Zou:2024osb} to Bumpy-AK waveform~\cite{Barack:2006pq}, analyzing $0.5$-year, SNR $50$ injections buried in simulated LISA noises. 

We find that the involvement of $\delta \tilde{Q}$ in Bumpy-AK waveform significantly changes the qualitative morphology of the likelihood surface compared with the AK waveform. It induces tens of thousands of points in parameter space where the likelihood exceeds either the value at the injected parameters or certain threshold value. We call the former supra-injection points (SIPs) 
and the latter supra-threshold points (STPs). As a result, a search focused on the unique global optimum of the likelihood, while reliable in the AK case with its far fewer SIPs and STPs, incurs substantial 
errors in the Bumpy-AK case. In particular, we observe that convergence to SIPs or STPs instead of the global maximum causes a significant bias in the estimation of critical parameters. A key result of this paper is that adopting an ensemble-based approach, where parameter estimates are derived from averaging across SIP and STP populations, rather than focusing on the likelihood at a single point, can substantially mitigate these biases while increasing the variance marginally. 
Our findings also illuminate the degeneracy pattern in Bumpy-AK parameter space 
which can be of value to future studies for probing new physics via EMRI data analysis.
	
This paper is structured as follows. Sec.~\ref{data} outlines the data model (TDI, noise, waveform). Sec.~\ref{method} details our methodology (reduced-dimensionality likelihood, PSO, nested optimization,  definition and methodology about SIPs and STPs and code implementation as well as computational costs). Results and discussion follow in Sec.~\ref{Results} and~\ref{Discussion}, respectively. 
	\section{data description}
	\label{data}
	This section outlines the core components of our analysis framework. First, we describe the TDI model, a critical method for suppressing laser frequency noise in space-borne GW detectors. Next, we present the theoretical noise model, including detector noise power spectral densities (PSDs) and key detection statistics of SNR, which are used to quantify sensitivity and template matching. Finally, we depict the Bumpy-AK parameterization, a framework for quantifying deviations from the Kerr metric to test GR in EMRI system.
\subsection{TDI}
Our TDI conventions adhere to the standards established in LDC~\cite{LDC-code-maunal}. The spaced-borne GW detector consists of three spacecrafts forming a triangle constellation where a single-link trip for laser involves two spacecrafts and their link, labeled by $\{s,l,r\}$ denoting the sender, link, and receiver, respectively. The unit arm vector for link $l$, denoted $\widehat{n}_l$, can be derived from the orbital motion of GW detectors. Given polarized waveform $h_{+,\times}$,  polarization tensors $\epsilon^{+,\times}$ and the polarization angle $\psi$,  the projection of GW strain on $\widehat{n}_l$, denoted $H_l$, is expressed as:
	\bea
	\begin{bmatrix}F_l^+ \\ F_l^{\times} \end{bmatrix}=&\begin{bmatrix} \cos(2\psi) & -\sin(2\psi)\\ \sin(2\psi)& \cos(2\psi)\end{bmatrix}\begin{bmatrix} (\widehat{n}_l \otimes \widehat{n}_l):\epsilon^{+}\\(\widehat{n}_l \otimes \widehat{n}_l):\epsilon^{\times}\\ \end{bmatrix}\;, \nonumber \\   
	H_l =& F_l^{+}h_{+}+F_l^{\times}h_{\times}\;, \label{eq:H_l}  \\
	\nonumber 
	\eea
where $F_l^{+, \times}$ are antenna pattern functions for link $l$, encoding the angular sensitivity to GWs. The contraction operator is defined for arbitrary tensors $U$ and $V$ as $U:V = \sum_{i,j} U_{ij}V_{ij}$. 
The outer product $\otimes$ acts on vectors $a$ and $b$ to produce a tensor $(a \otimes b)_{ij}=a_i b_j$.

The GW-induced single-link ($l$) strain response $y^{\rm GW}_{slr}$, is the difference between two measurements of $H_l$ at $s$ and $r$, respectively, expressed as:
	\bea
	y^{\rm GW}_{slr}(t) =& \frac{H_l(t-\hat{k}\cdot\widehat{R}_s-L_l) - H_l(t-\hat{k}\cdot\widehat{R}_r)}{2(1-\hat{k}\cdot\widehat{n}_l)}\;.
	\label{eq:TDI_y_slr}
	\eea 
The time delay and projection in $y^{\rm GW}_{slr}$ depend on the GW propagation vector $\hat{k}$, the orbital configuration of the detector (e.g., the position  $\widehat{R}_i$ of the $i$-th satellite for $i\in \{s,r\}$), the unit arm vector $\widehat{n}_l$ and the arm length $L_l$ of link $l$.
The link index $l$ corresponds to a cyclic permutation $1 \to 2 \to3 \to 1$ of the single-link labels $\{s,l,r\}$, with a positive sign assigned to this cyclic order and negative otherwise. 
	
Following the first-generation TDI adopted in LDC for EMRIs, the Michelson-type TDI channels $X$, $Y$, and $Z$ are constructed as follows:
	\bea
	X = &-2i\sin(\omega L)\left[e^{-i2\omega L}(y_{1-32}-y_{123}) + e^{-i\omega L}(y_{231}-y_{3-21}) \right]\;,  \nonumber \\
	Y=&-2i\sin(\omega L)\left[e^{-i2\omega L}(y_{2-13}-y_{231}) + e^{-i\omega L}(y_{312}-y_{1-32}) \right]\;, \nonumber \\
	Z=&-2i\sin(\omega L)\left[e^{-i2\omega L}(y_{3-21}-y_{312}) + e^{-i\omega L}(y_{123}-y_{2-13}) \right]\;. \label{eq:TDI-1} \\ \nonumber  
	\eea
The arm lengths $L_l$ are approximated as equal constants $L$ in the first-generation TDI, and the GW frequency $\omega$ is defined as the time derivative of the GW phase: $\omega=\frac{\partial \Phi^{\rm GW}}{\partial t}$. With the Michelson channels $X$, $Y$, $Z$ computed, we synthesize the noise-orthogonal TDI channels $A$, $E$, and $T$ via the transformation:
	\bea
	A =& \frac{Z-X}{\sqrt{2}}\;, \nonumber \\ 
	E =& \frac{X-2Y+Z}{\sqrt{6}}\;, \nonumber\\ 
	T =& \frac{X+Y+Z}{\sqrt{3}}\;. \label{eq:TDI-2}\\ 
	\nonumber
	\eea
Since the $T$ channel is insensitive to GW signals, this study utilizes only the $A$ and $E$ channels of GWs by taking the strain response $y^{\rm GW}_{slr}$ to Eq.~(\ref{eq:TDI-1}) and Eq.~(\ref{eq:TDI-2}).
\subsection{Noise model}
We adopt the theoretical PSD model for LISA, as formulated in~\cite{LDC-code-maunal}. Since the PSD of $A$ and $E$ channel are indentical, we adopt the unified notation $S_n(f)$ for their noise spectra in subsequent analyses, which is expressed as:
	\bea
	\label{eq:PSD}
	S_n^{A}(f) = S_n^{E}(f) = S_n(f)= 8\sin^2\omega L\big[4(1+\cos\omega L+\cos^2\omega L)S^{\rm Acc} +(2+\cos\omega L)S^{\rm IMS}\big]\;.
	\eea 
Here, $f$ denotes the Fourier frequency, $\omega = 2\pi f$ is the angular frequency, and $L$ represents the detector arm length, assumed constant under first-generation TDI conventions. Adopting the LISA reference noise model "SciRDv1"~\cite{LDC-code-maunal}, the acceleration noise $S^{\rm Acc}$
and instrumentation optical metrology system (IMS) noise $S^{\rm IMS}$ are defined as:
	\bes
	\begin{align}
		S^{\rm Acc}(f)=& \frac{9.0\times10^{-30}}{(2\pi fc)^2}\big[1+\big(\frac{0.4 {\rm mHz}}{f}\big)^2\big]\big[1+\big(\frac{f}{8{\rm mHz}}\big)^4\big]\frac{1}{\text{Hz}}\;,\\
		S^{\rm IMS}(f)=&2.25\times 10^{-22}\big(\frac{2\pi f}{c}\big)^2\big[1+(\frac{2 {\rm mHz}}{f})^4\big]\frac{1}{\text{Hz}}\;,\\ \nonumber
		\label{eq:LDC_noise_model}  
	\end{align}
	\ees 
where $c$ denotes the speed of light. Using the PSD, the noise-weighted inner product between two signals $\overline{a}$ and $\overline{b}$ is defined as:
	\bea
	(\overline{a}|\overline{b}) = \frac{1}{Nf_s}\sum_{k=0}^{N-1}\frac{\widetilde{a}_k\widetilde{b}^\ast_k + \widetilde{a}^\ast_k\widetilde{b}_k}{S_n(f_k)}\;,
	\label{eq:innerproduct}  
	\eea
where $\widetilde{a}$ denotes the discrete Fourier transform (DFT) of the time-domain signal  $\overline{a}$, and $\widetilde{a}^{\ast}$ is its complex conjugate (with analogous notation applying to  $\widetilde{b}$). The sampling frequency $f_s$, signal length $N$, and discrete frequency bins $f_k = kf_s/N$, $k = 0,1,\ldots,N-1$ define the spectral resolution of the DFT. 
For the noise-orthogonal TDI channels $\{A,E\}$ of the signals $\overline{a}$, the combined SNR are derived from the inner product structure as follows:
	\bea
	   {\rm SNR}^2_a =& (\overline{a}^A|\overline{a}^A)+(\overline{a}^E|\overline{a}^E)\;,\\
       \nonumber
     \eea  
	where the signal components in one channel manifest as null measurements for corresponding individual quantities (with analogous notation applying to  $\widetilde{b}$).
\subsection{Bumpy-AK Parameterization}
	\label{Bumpy-AK}
	The Bumpy-AK parameterization introduces a minimal deviation from the kerr metric by linearly perturbing its quadrupole moment~\cite{Barack:2006pq}. The dimensionless deviation parameter $\delta \tilde{Q}$ quantifies this offset from the Kerr value $\tilde{Q}_{\rm kerr}$. The modified multipole moment is subsequently formulated as:
	\be
	\tilde{Q} = \tilde{Q}_{\rm kerr}  + \delta \tilde{Q},
	\ee
	where $ \tilde{Q}_{\rm kerr}=-(S/M^2)^2$. As a result, the Bumpy-AK parameter space becomes $15$-dimensional, which can be partitioned into three distinct subsets:
	\bea
	\theta_1=&\{\mu, M, \lambda, S/M^2, e_0, \nu_0, \delta \tilde{Q}\} \;, \nonumber \\
	\theta_2=&\{\theta_s,\phi_s\}  \;, \nonumber \\ 
	\theta_3=&\{\theta_k, \phi_k, \phi_0, \tilde{\gamma_0}, \alpha_0, D\} \;, \nonumber \\ 
	\Theta =& \theta_1 \cup \theta_2 \cup \theta_3 \;.
	\eea
	The first subset, $\theta_1$, governs the orbital evolution of EMRIs with phase-coupled properties as discussed in Sec.~\ref{intro}, including the CO mass $\mu$, the MBH mass $M$, the inclination angle $\lambda$ between the CO's orbital angular momentum and the MBH's spin axis, the dimensionless MBH spin magnitude $S/M^2$, the initial eccentricity $e_0$ and orbital frequency $\nu_0$. The second subset, $\theta_2$, encodes the source’s sky location with ecliptic latitude $\theta_s$ and ecliptic longitude $\phi_s$, which introduces time-dependent modulations to the TDI channels through the detector’s orbital motion, analogous to the dynamical effects as $\theta_1$ but without comparable sensitivity. The third subset, $\theta_3$, contains parameters such as initial phases $\phi_0$ (orbital motion), $\tilde{\gamma_0}$ (pericenter precession), $\alpha_0$ (Lense-Thirring precession), the direction of MBH’s spin with ecliptic latitude $\theta_k$, ecliptic longitude $\phi_k$ and the distance $D$ from the GW source to the Solar System Barycenter, resulting in constant amplitude and phase modulations for each EMRI harmonic, which do not influence the time-varying components of the TDI channels but are critical for the phase coherence and amplitude scaling of the waveform with superposition of harmonics.
	
	The orbital dynamics of the system, governed by the parameter subset $\theta_1$, are modeled by the following ordinary differential equations (ODEs)~\cite{Barack:2006pq}:
	\bea
	\left(\frac{d\Phi}{dt}\right) &=& 2\pi\nu \;, \nonumber \\ 
	%
	\left(\frac{d\nu}{dt}\right) &=&
	\frac{96}{10\pi}(\mu/M^3)(2\pi M\nu)^{11/3}(1-e^2)^{-9/2}
	\bigl\{
	\left[1+(73/24)e^2+(37/96)e^4\right](1-e^2) \label{nudotKerr}\nonumber \\
	&&+ (2\pi M\nu)^{2/3}\left[(1273/336)-(2561/224)e^2-(3885/128)e^4
	-(13147/5376)e^6 \right] \nonumber \\
	&&- (2\pi M\nu)(S/M^2)\cos\lambda (1-e^2)^{-1/2}\bigl[(73/12)
	+ (1211/24)e^2
	+(3143/96)e^4 +(65/64)e^6 \bigr] \nonumber\\
	&&-(2\pi M\nu)^{4/3}\tilde{Q}(1-e^2)^{-1}
	\left[(33/16)+(359/32)e^2-(527/96)\sin^2\lambda\right]
	\bigr\}\;,  \nonumber\\ 
	%
	\left(\frac{d\tilde\gamma}{dt}\right) &=& 6\pi\nu(2\pi\nu M)^{2/3} (1-e^2)^{-1}
	\left[1+\frac{1}{4}(2\pi\nu M)^{2/3} (1-e^2)^{-1}(26-15e^2)\right] \nonumber \\
	&&-12\pi\nu\cos\lambda (S/M^2) (2\pi M\nu)(1-e^2)^{-3/2} \nonumber\\
	&& -\frac{3}{2}\pi\nu \tilde{Q}(2\pi M\nu)^{4/3}(1-e^2)^{-2}\left(5\cos\lambda-1\right)\;, \nonumber \\ 
	%
	\left(\frac{de}{dt}\right)  &=& -\frac{e}{15}(\mu/M^2) (1-e^2)^{-7/2} (2\pi M\nu)^{8/3}
	\bigl[(304+121e^2)(1-e^2)\bigl(1 + 12 (2\pi M\nu)^{2/3}\bigr) \, \nonumber \\
	&&- \frac{1}{56}(2\pi M\nu)^{2/3}\bigl( (8)(16705) + (12)(9082)e^2 - 25211e^4 \bigr)\bigr]\,
	\nonumber \\
	&&+ e (\mu/M^2)(S/M^2)\cos\lambda\,(2\pi M\nu)^{11/3}(1-e^2)^{-4}
	\, \bigl[(1364/5) + (5032/15)e^2 + (263/10)e^4\bigr]\;,  \nonumber\\ 
	%
	\left(\frac{d\alpha}{dt}\right)&=& 4\pi\nu (S/M^2) (2\pi M\nu)(1-e^2)^{-3/2}
	+3\pi\nu\tilde{Q}(2\pi M\nu)^{4/3}(1-e^2)^{-2}\cos\lambda\;. 
	\label{eq:ODEs}
	\eea
	The term $\tilde{Q}$ encapsulates higher-order corrections to the orbital dynamics of AK. The system of ODEs is numerically integrated using a fifth-order Cash-Karp embedded Runge-Kutta solver~\cite{NR_f90_v2}, ultilizing in fixed step-size for accuracy and efficiency. Once the orbital quantities are computed, the polarized gravitational waveform $h_{+,\times}$ is synthesized as a superposition of harmonic modes $(n,m)$, expressed as:
	\bea
	\overline{h}^{nm}_{+,\times}(\Theta) &= \sum_{n,m} \frac{1}{D} A^{nm}_{+,\times}(\theta_2,\lambda,\theta_k,\phi_k,e(t),\nu(t))e^{i\Phi^{nm}_0(\phi_0, \tilde{\gamma_0}, \alpha_0 )}e^{i\Phi^{nm}_t(\theta_1)}\;.
	\label{eq-hphc}
	\eea
	The phase of the $(n,m)$ harmonic mode is given by $\Phi^{nm}=n\phi(t) + 2\tilde{\gamma}(t) + m\alpha(t)$, where $\phi$ represents the orbital phase,  $\tilde{\gamma}$ accounts for pericenter precession, and $\alpha$ denotes the Lense-Thirring precession. The corresponding initial phase is $\Phi^{nm}_0=n\phi_0 + 2\tilde{\gamma}_0 + m\alpha_0$. The prefactor of $2$ in $\tilde{\gamma}$ reflects the quadrupolar-dominated nature of the AK waveform model, which assumes that leading-order gravitational radiation is sourced by the mass quadrupole moment.
	
	In LDC, the AK waveforms incorporate harmonics with indices $n\in \{1,2,3,4,5\}$ and azimuthal modes $|m|\le 2$ per $n$, yielding a total of $25$ harmonics. As demonstrated in Table~\ref{tab-1-top-10-harmonics}, the top-$10$ harmonics consistently capture $\ge 90\%$ of the total SNR for systems with low eccentricity and moderate mass ratios, aligning with trends observed for the astrophysically motivated default EMRI parameters whose values are adopted from Ref.~\cite{Barack:2006pq}. For moderate eccentricities ($e_0 \le0.3$), harmonics with $n=2$ and $n=3$ dominate over other $n$, thus comprising the majority of the top-$10$ SNR-contributing harmonics. To balance accuracy and computational cost, our analysis focuses exclusively on these dominant top-$10$ harmonics. A comprehensive discussion of harmonic selection criteria, SNR fraction trade-offs effects are provided in~\cite{Zou:2024jqv,Zou:2024osb}.
	\begin{table}[htb]
		\centering
		\begin{tabular}{|c|c|c|c|c|c|}
			\hline
			\makecell{Descending order \\ SNR}   &\makecell{Default\\ parameters} &$M=10^{5.5} M_{\odot}$ &\makecell{$M=10^{6.5} M_{\odot}$\\ $\nu_0=0.2723873631$ mHz}  & $e_0=0.3$ & $e_0=0.4$ \\
			\hline
			$1$ &$2$/$0.617$ &$2$/$0.503$ &$2$/$0.411$ &$2$/$0.442$ &$4$/$0.36$   \\
			$2$ &$3$/$0.307$ &$3$/$0.382$ &$3$/$0.386$ &$3$/$0.383$ &$5$/$0.333$  \\
			{\rm \textbf{Power/SNR~fraction}} 
            &$\textbf{0.924}$/$\textbf{0.961}$
            &$\textbf{0.885}$/$\textbf{0.941}$ 
            &$\textbf{0.797}$/$\textbf{0.893}$  
            &$\textbf{0.825}$/$\textbf{0.908}$ 
            &$\textbf{0.693}$/$\textbf{0.833}$   \\   
			$3$ &$4$/$0.062$ &$4$/$0.098$ &$4$/$0.157$ &$4$/$0.134$ &$3$/$0.246$   \\
			$4$ &$5$/$0.009$ &$$5$/0.016$ &$5$/$0.043$ &$5$/$0.033$ &$2$/$0.06$  \\
			$5$ &$1$/$0.005$ &$1$/$0.002$ &$1$/$0.002$ &$1$/$0.007$ &$1$/$0.001$  \\
			\hline
		\end{tabular}
		\caption{Illustration of the harmonic SNR rank in Bumpy-AK waveform for different configurations. The header row identifies parameters altered from their default values to create distinct source configurations, while other parameters remain fixed. Table entries follow the $n/y$ format, where $n$ is the harmonic index and $y$ is its fractional power contribution. The $1$st column denotes the descending order in terms of SNR for harmonics $n$,  the ${\rm Power/SNR~fraction}$ row in bold quantifies the cumulative power and aggregate SNR contribution of the top-$10$ harmonics, emphasizing their collective dominance in the waveform.  
		The default parameters are listed in Table~\ref{tab-2-injection}, defining the baseline configuration used to generate Bumpy-AK waveforms for all subsequent tests of non-Kerr spacetime signatures.
		}\label{tab-1-top-10-harmonics}
	\end{table}

	\section{Methodology}
	\label{method}
	In this section, firstly we present the general $15$-dimensional likelihood function for Bumpy-AK waveform and its sharp $7$-dimensional peak structure, and construct a variable-separated likelihood function via linear waveform decomposition, decoupling the parameter subset $(\theta_1, \theta_2)$ from $\theta_3$ based on their distinct roles in shaping the time evolution of the TDI channels. Then we describe the PSO algorithm and its advantages for searching AK and Bumpy-AK signals. Next we introduce the nested optimization strategy to address the high dimensionality where a global optimizer targets the parameters in $\{\theta_1, \theta_2\}$ and a local optimizer refines the parameters in $\theta_3$, which can balances exploration and precision for searching both AK and Bumpy-AK signals. Subsequently, we formulate the definitions of SIP and STP, alongside ensemble-based statistical frameworks validated for mitigating systematic parameter estimation errors.  Finally, we detail our computational approximations and parallelized framework tailored to enable efficient searches.
	
	\subsection{Generalized Likelihood Ratio Test}
	\label{LLR} 
\begin{figure}[htbp]
		\centering 
	\includegraphics[height=4.0in]{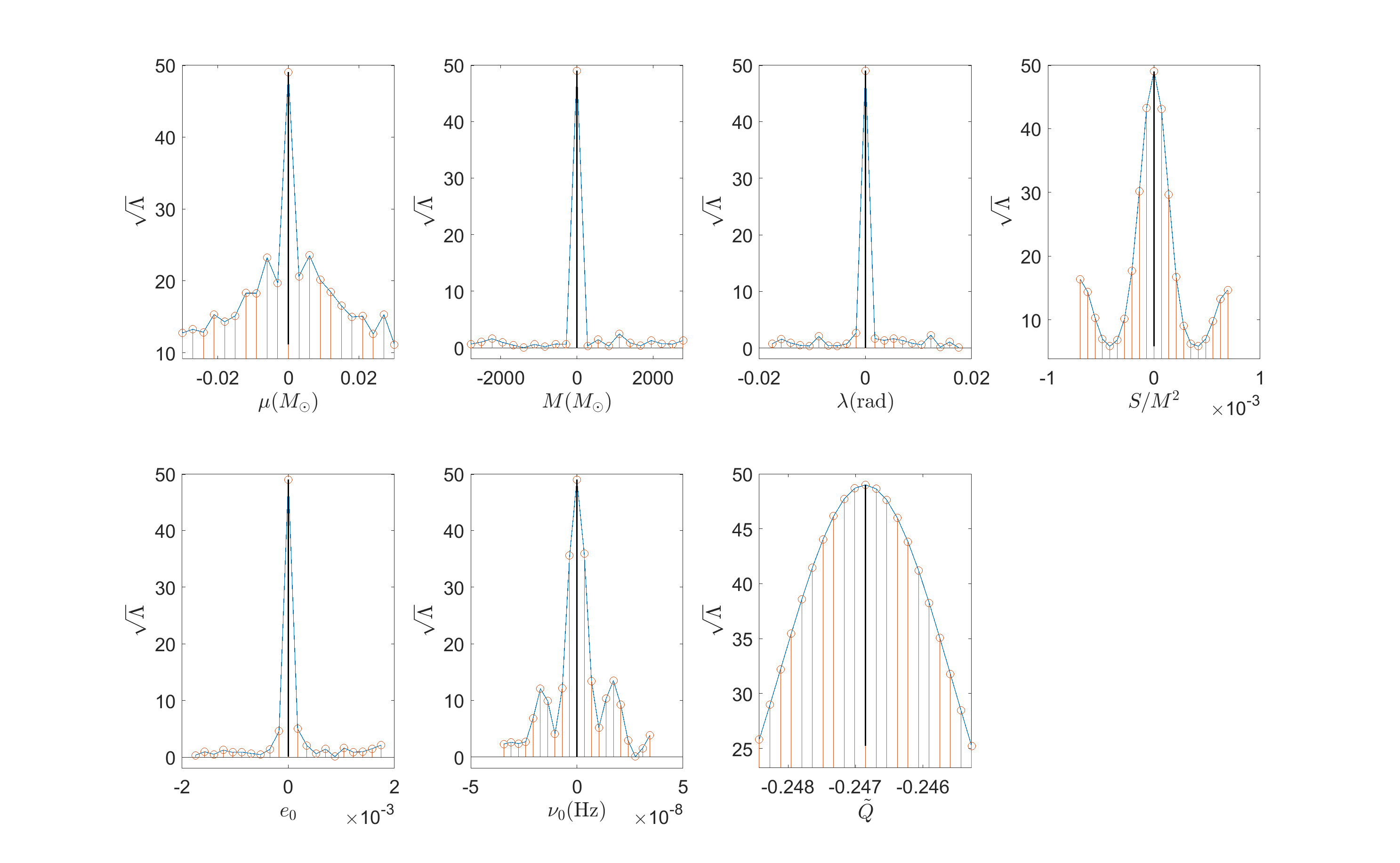}
		\caption{Illustration of the $7$-dimensional sharp peak of the signal morphology where the x-axis is the distance away from the injected value as given in table~\ref{tab-2-injection} for each parameter in set $\theta_1$, and $\tilde{Q}_{\rm kerr}=-0.25, \delta \tilde{Q} = 3.1\times 10^{-3}$. The other parameters are fixed when each parameter is varying.}
		\label{fig-7D-sharp-peak}  
	\end{figure}

	For TDI channels $I\in \{A,E\}$, let $\overline{d}^I$ represents the observed data stream containing an embedded GW signal $\overline{h}^I(\Theta)$, thus the data model is $\overline{d}^I$= $\overline{h}^I(\Theta)$+ $\overline{n}^I$, where $\Theta$ denotes the full $15$-dimensional parameter sets and $\overline{n}^I$ represents the instrumental noise with its PSD given in Eq.~(\ref{eq:PSD}).  The log-likelihood ratio (LLR)~\cite{Kay:GLRT}, a key detection statistic, is expressed as:
	\bea
	\Lambda(\Theta) = \sum_{I\in \{A, E\}} \left[-(\overline{h}^I(\Theta)|\overline{h}^I(\Theta)) + 2(\overline{d}^I|\overline{h}^I(\Theta))\right]\;.
	\label{eq:14D_LLR}
	\eea 
	Given the LLR function $\Lambda(\Theta)$, the goal of data analysis is twofold: (1) to identify the global maximum likelihood estimate $\widehat{\Theta}  =  \argmax\limits_{\Theta} \Lambda(\Theta)$, corresponding to the best-fit parameters, or (2) to sample its posterior probability density function (PDF) via Bayesian inference. As demonstrated in Fig.~\ref{fig-7D-sharp-peak}, the estimated $\widehat{\Theta}$ resides near the injected values of the $7$ phase-coupled parameters, with deviations arising from noise fluctuations.  Crucially, the subset $\theta_1$ generates a sharply localized 7-dimensional peak in $\Lambda(\Theta)$, consistent with the phase sensitivity discussed in Sec.~\ref{intro}. The union $\{\theta_1 \cup \theta_2\}$ govern the time-varying effects (e.g., orbital dynamics, Doppler shifts, antenna pattern evolution) in TDI channels, whereas the $\theta_3$ affects static waveform properties. 
	The divergent roles of $\theta_1$, $\theta_2$ and $\theta_3$ in shaping the needle-in-haystack signal morphology are illustrated through comparative analysis of Fig.~1 in~\cite{Xiaobo:2025jkw} and Fig.~\ref{fig-7D-sharp-peak} in this study, which render the full $15$-dimensional likelihood optimization computationally inefficient. To address this challenge, we develop a variable-separation likelihood maximization procedure, which decouples the $\theta_3$ from $\{\theta_1 \cup \theta_2\}$  via waveform decomposition. The methodology proceeds as follows.
	
	We start from analytically maximizing the distance $D=1/A$ by $\frac{\partial  \Lambda(\Theta)}{\partial A} = 0$~\cite{Babak:2009ua} where $\overline{h}^I(\Theta)=A\overline{s}^I(\theta^\prime)$ and $\theta^\prime = \Theta \setminus \{A\}$. $\theta^\prime$ is the $14$-dimensional parameter set,  $\overline{s}^I(\theta^\prime)$ is the $14$-dimensional GW signal. By doing that, we obtain that 
	\bea
		\widehat{\mathrm{A}} =& \argmax\limits_{\mathrm{A}}\Lambda(\Theta) = \frac{\big[\sum_{I\in \{A, E\}} (\overline{d}^I|\overline{s}^I(\theta^\prime))\big]}{\big[\sum_{I\in \{A, E\}} (\overline{s}^I(\theta^\prime)|\overline{s}^I(\theta^\prime))\big]}\;,
	\eea
	leads to a $14$-dimensional likelihood 
	\bea
	\rho^2(\theta^\prime) & = \Lambda(\widehat{\mathrm{A}},\theta^\prime)= \frac{\big[\sum_{I\in \{A, E\}} (\overline{d}^I|\overline{s}^I(\theta^\prime))\big]^2}{\big[\sum_{I\in \{A, E\}} (\overline{s}^I(\theta^\prime)|\overline{s}^I(\theta^\prime))\big]}\;.
	\label{eq:glrt1}
	\eea 
	Here the goal turns to $L_G = \max\limits_{\theta^\prime} \rho^2(\theta^\prime)$. 
	
	We next conduct a linear decomposition of the $14$-dimensional polarized waveform $\overline{s}^{nm}_{+,\times}(\theta^\prime)$~\cite{Zou:2024osb}, structured as follows:
	\bea
	\overline{s}^{nm}_{+,\times}(\theta^\prime) =& \sum_{n,m} A^{c,m}_{+,\times}(\theta_2,\lambda,\theta_k,\phi_k)e^{i\Phi_0^{nm}(\phi_0, \tilde{\gamma_0}, \alpha_0)}\left[A^{t,n}(e(t),\mu(t))e^{i\Phi^{nm}_t(\theta_1)}\right]\;.
	\label{eq:hphc_split}    
	\\ \nonumber
	\eea
	Here the amplitudes of the polarized waveform $\overline{s}^{nm}_{+,\times}(\theta^\prime)$ are split into the time independent $A^{c,m}_{+,\times}$ and time dependent $A^{t,n}$, labeled with superscript $c$ (constant) and $t$ (time-dependent) respectively. Let  $\theta^{\prime\prime}$ denotes the union of phase-coupled and sky position parameters, $\theta^{\prime\prime}=\{\theta_1 \cup \theta_2\}$, and define the time-dependent harmonic amplitude-phase term as: $\overline{x}^{nm}(\theta_1)=A^{t,n}(e(t),\mu(t))e^{i\Phi^{nm}_t(\theta_1)}$. Henceforth, we adopt the single index $i$ to label harmonics in place of the double-index notation $(n,m)$. By combining Eq.~(\ref{eq:hphc_split}) with the single-link antenna pattern functions $F_l^{+,\times}$, we construct the $14$-dimensional GW projection on $\widehat{n}_l$, denoted $\overline{s}^{i}_{l}(\theta^\prime)$, expressed as:
	\be
	\label{eq:8D_decomposition_hphc}
	\begin{split}
		\overline{s}^{i}_{l}(\theta^\prime)=&F_l^{+}(\theta_2,\psi){\rm Re}(\overline{s}^{i}_{+}(\theta^\prime))+F_l^{\times}(\theta_2,\psi){\rm Re}(\overline{s}^{i}_{\times}(\theta^\prime))\;,\\ 
		=&{\rm Re}(e^{i\Phi^i_0}A^c_{+}(\theta_k,\phi_k,\theta_2,\lambda))F_l^{+}(\theta_2,\psi){\rm Re}(\overline{x}^i(\theta_1))  \\ 
		-&{\rm Im}(e^{i\Phi^i_0}A^c_{+}(\theta_k,\phi_k,\theta_2,\lambda))F_l^{+}(\theta_2,\psi){\rm Im}(\overline{x}^i(\theta_1))  \\ 
		+&{\rm Re}(e^{i\Phi^i_0}A^c_{\times}(\theta_k,\phi_k,\theta_2,\lambda))F_l^{\times}(\theta_2,\psi){\rm Re}(\overline{x}^i(\theta_1))  \\ 
		-&{\rm Im}(e^{i\Phi^i_0}A^c_{\times}(\theta_k,\phi_k,\theta_2,\lambda))F_l^{\times}(\theta_2,\psi){\rm Im}(\overline{x}^i(\theta_1)), \\ 
		=&\sum_{p=1}^{4}a^{i}_{p} \overline{x}^{i}_{l,p}(\theta^{\prime\prime})\;. \\
	\end{split}
	\ee
	Eq.~(\ref{eq:8D_decomposition_hphc}) reveals that all time-dependent components involving the time-varying orbital and detector motion effects are encapsulated within the linearly decomposed term $\overline{x}^{i}_{l,p}(\theta^{\prime\prime})$. Furthermore, the two different TDI delays are projected to $H_l$ in Eq.~(\ref{eq:H_l}) (corresponding to $\overline{s}^{i}_{l}(\theta^\prime)$ here) respectively, and the two delayed $H_l$ share the same linear decomposition as Eq.~(\ref{eq:8D_decomposition_hphc}). Consequently, the linear decomposition can be preserved from $H_l$ to single-link strain response $y^{\rm GW}_{slr}$ in Eq.~(\ref{eq:TDI_y_slr}), and further to TDI channels $X$, $Y$, $Z$ and  $A$, $E$ because of the linear equation forms in Eq.~(\ref{eq:TDI-1}) and Eq.~(\ref{eq:TDI-2}). As a result, the $14$-dimensional TDI response  $\overline{s}^{I,i}(\theta^\prime)$ is expressed through the following structure:
	\bea
	\overline{s}^{I,i}(\theta^\prime)&=&\sum_{p=1}^{4}a^{i}_{p} \overline{x}^{I,i}_{p}(\theta^{\prime\prime})\;. 
	\label{eq:8D_decomposition_TDI}\\ \nonumber
	\eea
	By applying the linear decomposition from Eq.~(\ref{eq:8D_decomposition_TDI}) to the inner products in Eq.~(\ref{eq:glrt1}), we express the inner products in terms of the decomposed components as follows:
	\bea
	(\overline{d}^I(\theta^\prime)|\overline{s}^I(\theta^\prime))=\sum_{i=1}^{N}\sum_{p=1}^4 a^i_p (\overline{d}^I\big|\overline{x}^{I,i}_{p}(\theta^{\prime\prime}))\;,
	\nonumber \\  
	(\overline{s}^I(\theta^\prime)|\overline{s}^I(\theta^\prime))=\sum_{i=1}^{N}\sum_{j=1}^{N}\sum_{p=1}^4\sum_{q=1}^4 a^i_p a^j_q (\overline{x}^{I,i}_{p}(\theta^{\prime\prime}) \big| \overline{x}^{I,j}_{q}(\theta^{\prime\prime}))\;, 
	\label{eq:13D_innerproduct_full}
	\eea
	where $N$ is the number of harnonics involved, and $N=10$ is used in this paper as discussed in Sec.~\ref{Bumpy-AK}. 
	
	Thus far, we have achieved variable separation between the parameter subsets $\theta^{\prime\prime}$
	(phase-coupled and sky position parameters) and $\theta_3$
	(static phase/amplitude constants) within the detection statistics defined in Eq.~(\ref{eq:glrt1}). This separation enables a reduced-dimensionality LLR framework via nested optimization~\cite{Zou:2024osb} where $\theta_3$ are iteratively optimized in distinct stages. The methodology is elaborated in subsequent sections.
	\subsection{Particle Swarm Optimization}
	\label{PSO}
	PSO is a metaheuristic optimization technique~\cite{Kennedy:1995,Kennedy:2007,Mohanty_book_2018} particularly effective for high-dimensional, multimodal parameter spaces inherent to GW astronomy.
	For a fitness function $f(\overline{x})$ with input vector $\overline{x}$, the algorithm employs $N_p$ particles that iteratively explore the search space of $\overline{x}$, balancing global exploration and local exploitation to converge on the optimal position $\overline{x}_{*}$, which satisfies $f(\overline{x}_{*})>f(\overline{x}), \forall \overline{x}$ in the search space. The position and velocity updates are governed by:
	\bea
	\overline{v}_{i}(t+1) & = & w \overline{v}_{i}(t) + c_1 r_1 (\overline{p}_{i}(t) - \overline{x}_{i}(t)) + c_2 r_2 (\overline{g}(t) - \overline{x}_{i}(t))\;.
	\label{eq:velocity_update}  \\
	\overline{x}_{i}(t+1) & = & \overline{x}_{i}(t) + \overline{v}_{i}(t+1)\;, \label{eq:position_update}\\
	\nonumber 
	\eea
	where $\overline{x}_i(t)$ and $\overline{v}_i(t)$ denotes the position and velocity vector of the $i$-th particle at iteration $t$, respectively. The personal best ({\it pbest}) of $\overline{p}_{i}$ and global best ({\it gbest}) of $\overline{g}$ represent the position vector for historical optimum 
    of $i$-th particle and the collective optimum of the swarm, respectively. 
    Through successive updates, these particles collectively converge toward the global optimum $\overline{x}_{*}$, with the process terminating once the maximum iteration  $N_{\rm iter}$ is reached. 
	The inertia weight $w$ is initially set to $0.9$ and linearly decreasing to $0.4$, moderates the balance between exploration and exploitation. Acceleration coefficients $c_1$, $c_2$ ($c_1=c_2=2$) scale the influence of the cognitive and social terms respectively, and $r_{1,2}$ are independent random numbers distributed uniformly in $[0,1]$ for each component which provide stochasticity. Velocities are clamped to $|v_i^j| \le V_{\rm max}$, where $V_{\rm max}$ is usually chosen from the range $[0.2, 0.9]$, where its larger values enhance exploratory capabilities.
	
	For EMRI-related applications, we implement a \textbf{local best}  ({\it lbest}) PSO variant~\cite{Kennedy:2007}, replacing $\overline{g}(t)$ in Eq.~(\ref{eq:velocity_update}) with a neighborhood-optimal position $\overline{p}_{\text{local}, i}(t)$  chosen by:
	\bea
	f(\overline{p}_{\text{local}, i}(t)) = \max\limits_{j\in \mathcal{N}_i} f(\overline{p}_j(t))\;,
	\eea
	where $\mathcal{N}_i$ defines a ring topology ($\mathcal{N}_i=\{i-1, i, i+1\}, i \in 1,2,3,...,N_p$ in which the first and last particle are circularly connected) to enhance local exploitation. While computationally more demanding than standard PSO, this variant excels in resolving the sharp $7$-dimensional likelihood peaks (Fig.~\ref{fig-7D-sharp-peak}) and secondary maxima characteristic of the Bumpy-AK signals, where the LLR exhibits a needle-in-a-haystack morphology.
	
	Key parameters suitable for EMRI data analysis include $N_p=40$ particles and $N_{\rm iter}=10000$ iterations per run, with $N_{\rm run}=6$ independent trials to enhance robustness and 
    $V_{\rm max}=0.5$ based on our previous work~\cite{Zou:2024osb}. Success criteria require the best-fit LLR to exceed that of the injected signal~\cite{Wang:2010jma}.
	
	The computational efficiency of PSO is particularly advantageous for EMRI data analysis, where traditional MCMC methods struggle due to the challenges for sampling density required to resolve high-dimensional peaks. PSO achieves comparable accuracy with orders of magnitude fewer likelihood evaluations, as evidenced in LIGO's compact binary coalescence studies~\cite{Weerathunga:2017qce,Normandin:2018lzk,Normandin:2020ltl}. This efficiency, combined with nested optimization strategies, positions PSO as a critical tool for detecting deviations from kerr black holes in EMRI system in future space-borne GW missions.
	
	\subsection{Nested Optimization}
	As detailed in Sec.~\ref{LLR}, the parameter subset $\theta_3$ is decoupled from $\theta^{\prime\prime}$ in the LLR (Eq.~(\ref{eq:glrt1})) through linear decomposition of the inner product (Eq.~(\ref{eq:13D_innerproduct_full})). This separation leverages the simpler structure of the $\theta_3$ subspace, which lacks sharp peaks and exhibits minimal multimodality across its parameter range, as demonstrated in~\cite{Zou:2024osb}.  Building on this, we adopt a nested optimization strategy that partitions the high-dimensional search into two distinct tasks, i.e.,
	\bea
	\label{eq:LLR_9D}
	L_G = \max\limits_{\theta^{\prime\prime}}\max\limits_{\theta_3} \rho^2(\theta^{\prime\prime},\theta_3)\;,
	\eea
	a computationally challenging global optimization over $\theta^{\prime\prime}$ (governing the time dependent components in TDI channels) and a computationally tractable local optimization over $\theta_3$ (regulating the time independent components in TDI channels). The former is addressed using PSO, leveraging its robust global search capabilities, while the latter employs the Nelder-Mead simplex algorithm~\cite{Nelder-Mead,gsl_local_minimizer}, a gradient-free local optimization method with high cost efficiency for smooth, low-dimensional spaces.
	
\begin{figure}[htbp]
		\centering 
		\includegraphics[height=4.0in]{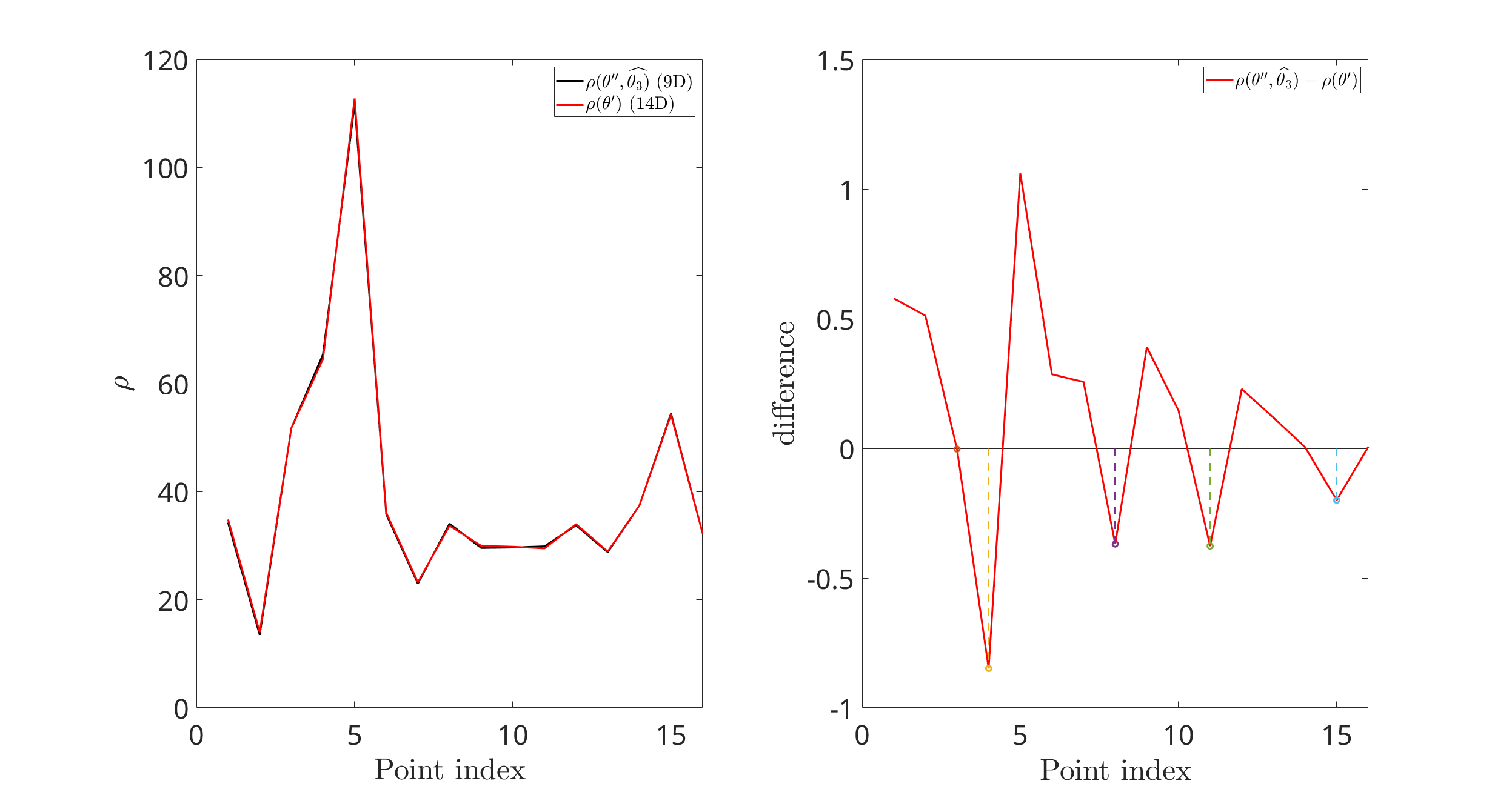}
		\caption{Illustration of the reduced dimensionality likelihoods for $16$ points randomly chosen in the parameter space with varying mass ratio, inclination, eccentricity and source sky location, where $\widehat{\theta_3}$ in $\rho(\theta^{\prime\prime},\widehat{\theta_3)}$ is the optimized solution returned by the local optimizer for each $\theta^{\prime\prime}$.} 
		\label{fig-LLR_9D_15D}  
	\end{figure}

	The reduced-dimensionality likelihood framework focuses on the $9$-dimensional $\theta^{\prime\prime}$ in Bumpy-AK waveform and derive its computational efficiency from precomputed  $\theta^{\prime\prime}$-dependent inner product terms in Eq.~(\ref{eq:13D_innerproduct_full}), which reduces the $\theta_3$ optimization to solving a polynomial function. Each local optimization over  $\theta_3$ requires only a fraction of milliseconds using a $2.3$GHz single-core processor. To ensure robustness, we initialize the Nelder-Mead algorithm with $243$ uniformly spaced grid points across the $5$ angular parameters in  $\theta_3$, sampling each angle at  $[0, 2\pi/3, 4\pi/3]$. This grid design ($3^5$ combinations) guarantees comprehensive coverage of global optima configurations in $\theta_3$ subspace for each $\theta^{\prime\prime}$ while maintaining computational efficiency, with total evaluation times under a few tens of milliseconds per PSO iteration. The utilization of the $9$-dimensional likelihood not only offers computational efficiency but also demonstrates a significant capacity to enhance LLR values, as illustrated in Fig.~\ref{fig-LLR_9D_15D}. This improvement is observed at $12$ out of $16$ random locations, where the $9$-dimensional likelihood effectively acts as a clustering mechanism that mitigates the LLR fluctuations originating from the $\theta_3$. For the remaining  $4$ locations, the marginal reduction in LLR values remains within acceptable limits, though the underlying causes warrant further investigation to clarify the behavior of the local optimizer in certain source configurations.
	
	The workflow begins with analytical maximization of the distance $D$ in the $15$-dimensional LLR in Eq.~(\ref{eq:14D_LLR}). The parameters in the subset $\theta^{\prime\prime}$ are then optimized globally via PSO, where each particle’s fitness is evaluated by solving the  $\theta_3$ subspace using Nelder-Mead as shown in Eq.~(\ref{eq:LLR_9D}). This nested approach achieves full parameter estimation by combining PSO’s global exploration with the efficiency of local gradient-free methods, effectively addressing the high dimensionality challenge posed by the $15$-dimensional parameter space and eliminating the secondary maxima resulting from $\theta_3$ in Bumpy-AK likelihood surfaces.

	\subsection{SIP and STP}\label{SIP-STP}    
	Unlike previous AK based EMRI analyses~\cite{Zou:2024jqv,Zou:2024osb} which yield sparse SIPs and STPs, the Bumpy-AK case produces tens of thousands of SIPs and STPs due to enhanced degeneracies in the $15$-dimensional parameter space. Critical to Bumpy-AK waveform analysis, SIPs and STPs are formally defined as:
        \bea
        \label{eq:SIP-STP}
\rho(\overline{x})>\rho(\overline{x}_{\rm true}) \;, \quad {\rm \overline{x} \in SIPs} \;, \nonumber\\
	\rho(\overline{x})>\rho_{\rm threshold}\;,  \quad {\rm \overline{x} \in STPs} \;,\\
	\nonumber
	\eea
    where $\rho$ is the LLR defined in Eq.~(\ref{eq:glrt1}), and both can be retrieved from the stored 
    history of 
    all PSO iterations for each PSO search. The predefined LLR threshold $\rho_{\rm threshold}$ is given by certain level of False alarm rate (FAR) for reliable GW detection. In effect, $\rho_{\rm threshold}$ is used to identify peaks in the LLR that caused by GWs by discarding peaks arising from noise alone.
	For benchmark datasets with known signal parameters $\overline{x}_{\rm true}$ (e.g., LDC-$1.2$~\cite{Baghi:2022ucj}), we identify the SIPs which are indistinguishable from the true signal location $\overline{x}_{\rm true}$ in terms of LLR values, demonstrating how degeneracies from $\delta \tilde{Q}$ dominate Bumpy-AK likelihood surfaces, validating the performance of their ensemble statistics rather than the optimal individual for reducing the errors in parameter estimation. Given this premise, we ascertain STPs through a certain FAR related LLR threshold originating from GWs rather than noise artifacts for blind analyses, alleviate the degeneracy from $\delta \tilde{Q}$ by performing the same ensemble statistics for STPs.
	
	In this paper, we study three different ways of estimating the parameters of a signal in a given data realization as follows:
	\begin{enumerate}
		\item \textbf{Standard parameter estimation:} The best solution (best {\it gbest} point across the $6$ independent searches) provides the point estimate of the injection parameters for each data realization. 
		\item  \textbf{SIP-based parameter estimation:}  For each data realization, compute the sample mean and standard deviation of its SIPs. Instead of using the best of {\it gbest} solutions from the $6$ PSO runs, the sample mean of all the SIPs found from the runs serves as the point estimate of the injection parameters. 
		\item \textbf{STP-based parameter estimation:} This scheme is identical to the SIP-based one except the SIPs are replaced by STPs.
	\end{enumerate} 
    It is important to emphasize here that the SIP-based parameters estimation method is purely for obtaining simulation-based insights. Obviously, it cannot be used for analyzing data where the true signal parameters are unknown. For this, one would use the STP-based parameter estimation method. 

	\subsection{Implementation and Computational Cost}
	\label{Implementation-cost}
	The reduced-dimensionality likelihood framework incurs non-trivial computational costs, particularly due to the calculations of TDI channels for each harmonic term in Eq.~(\ref{eq:8D_decomposition_TDI}). To mitigate these expenses, we implement a multi-tiered parallelization strategy using high-performance computing (HPC) resources.
	
	Two computational approximations
	\begin{enumerate}
		\item  \textbf{Orbital Dynamics Integration}:  Solving the coupled ODEs in Eq.~(\ref{eq:ODEs}) for prolonged inspiral durations is computationally intensive. Leveraging the slow adiabatic evolution of EMRI systems, we first integrate the ODEs at a coarse cadence of $15,360$ seconds, then interpolate the solutions to the LISA observational cadence of $15$ seconds. This reduces runtime significantly without sacrificing waveform accuracy too much.
		\item \textbf{Inner Product Approximation}: Following~\cite{Babak:2009ua}, we neglect cross-harmonic inner products in Eq.~(\ref{eq:13D_innerproduct_full}), which are orders of magnitude smaller than self-harmonic terms due to orthogonality. This approximation reduces computational complexity by a factor proportional to the number of harmonics.
	\end{enumerate}
	and two parallelization implementations
	\begin{enumerate}	 
		\item \textbf{Intra-Node Parallelism}: Using OpenMP~\cite{OpenMP}, we parallelize three key components across $64$ threads per node:
		polarized waveform generation in Eq.~(\ref{eq:8D_decomposition_hphc}),
		 calculations of TDI channels in Eq.~(\ref{eq:8D_decomposition_TDI}) and 
		inner product evaluations in Eq.~(\ref{eq:13D_innerproduct_full}).
		\item \textbf{Inter-Node Distribution}: Likelihood evaluations for $N_p=40$ particles are distributed across $40$ nodes via OpenMPI~\cite{MPI}, enabling concurrent exploration of the parameter space.
	\end{enumerate}
	are applied as mentioned above. The combined strategy enhances computational efficiency and optimizes resource utilization, as demonstrated in Table~\ref{Tab:cost}. Notably, the computational cost of approximately $0.4$ seconds per likelihood evaluation confirms its feasibility for large-scale EMRI data analyses.
	\begin{table}[htb]
		\centering
		\begin{tabular}{|c|c|c|c|c|}
			\hline
			\makecell{Cost of single \\Likelihood evaluation \\(cost of each PSO iteration)} & \makecell{Cost of single \\PSO search \\ $10000$ iterations} & \makecell{Cost of each data realization \\$6$ independent PSO search} & \makecell{Cost of each $30$ data realizations \\$180$ independent PSO search \\ (Case $1$,  Case $2$)} & \makecell{Totoal cost\\ Case $1\sim4$ } \\
			\hline
			$\sim0.4$ sec &\makecell{$\sim4000$ sec \\ $\sim 1.1$ hours} & $\sim 8$ hours & $\sim 240$ hours   & $\sim 500$ hours\\
			\hline
		\end{tabular}
		\begin{tabular}{|c|c|c|c|c|}
			\hline
          \makecell{Number of threads \\ used by OpenMP} & 
          \makecell{Number of nodes \\ used by OpenMPI} &
          CPU memory &
          CPU Clock Rate & 
          Total CPU hours
          \\
			\hline
			 $64$ & $40$ & $200$ GB & $1.5\sim 2.25$ GHz &$\sim 1.3\times 10^6$\\
			\hline
		\end{tabular}
		\caption{Demonstration of the computational cost for signal duration of $0.5$-year. The top panel presents the implementation costs, while the bottom panel provides the technical specifications of the HPC cluster employed in this study. }\label{Tab:cost}
	\end{table}
	This balanced approach with strategic approximations coupled with massive parallelism enables practical implementation of the nested optimization framework while maintaining physical fidelity.

	\section{Results}
	\label{Results}
	This section begins by specifying the injected signal's source parameters and corresponding search ranges. We then present four distinct test cases where PSO-based matched filtering are implemented using the reduced-dimensional likelihoods defined in Eq.~(\ref{eq:glrt1}) and Eq.~(\ref{eq:13D_innerproduct_full}). Each case varies the relationship between the injected signal and template waveform to assess robustness under different astrophysical or instrumental scenarios. Finally, we analyze and interpret the results of each case to evaluate the algorithm’s performance. 
	
	To balance the computational feasibility of resolving $7$-dimensional sharp peak in Bumpy-AK likelihood surfaces against search algorithm constraints, we adopt $0.5$-year duration signals with SNR $50$ as our benchmark injection. Using the pipeline, the following $4$ cases are designed to test its robustness for exploring the non-Kerr features and the effect of template uncertainty in EMRI system, 
	\begin{enumerate}
		\item  \textbf{Case $1$ (Baseline Validation):} A non-deviating signal ($\delta \tilde{Q}=0$) is injected to validate the pipeline’s ability to recover canonical (Kerr-like) parameters. This serves as a null test to confirm baseline functionality and calibration.
		\item \textbf{Case $2$ (Bumpy Parameter Constraints):} A signal with moderate deviations ( $\delta \tilde{Q}= 3.1 \times 10^{-3}$) is injected to quantify the pipeline’s parameter constraint capabilities. The chosen deviation corresponds to $4 \sigma$ in its FIM length, probing the pipeline’s performance for resolving non-Kerr features.
		
		\item \textbf{Case $3$ (Waveform Inaccuracy $\textrm{I}$-small):} To assess systematic biases from waveform modeling limitations, we simulate a small mismatch where the injected signal uses AK waveform, while the recovery template employs the Bumpy-AK template. This tests whether subdominant waveform uncertainties induce spurious detections of non-Kerr parameters.
		
		\item \textbf{Case $4$ (Waveform Inaccuracy $\textrm{II }$-large):} Repeating Case 3 with larger waveform discrepancies using Fast EMRI Waveforms (FEW)~\cite{Katz:2021yft} as injection, this evaluates degeneracies between astrophysical deviations ($\delta \tilde{Q}$) and waveform modeling errors. We quantify whether strong mismodeled dynamics mimic or obscure genuine non-Kerr signatures.
	\end{enumerate}
	\begin{table}[htb]
		\centering
		\begin{tabular}{|c|c|c|c|}
			\hline
			\diagbox{Parameters}{Values}   & Injected values & FIM $\sigma$  & \makecell{Search range width \\ in FIM $\sigma$ \\ except $\theta_s$, $\phi_s$}\\
			\hline
			$\mu(M_{\odot})$ & $10$  &$1.5057\times10^{-2}$ &$10$ \\		
			$M(M_{\odot})$ & $10^6$  &$1.3925\times10^{3}$  &$10$\\		
			$\lambda({\rm rad})$ & $\pi/3$  &$8.8003\times10^{-3}$  &$10$\\		
			$S/M^2$ & $0.5$  &$3.4710\times10^{-4}$  &$10$ \\		
			$e_0$ & $2.2468\times10^{-1}$  &$8.7545\times10^{-4}$  &$10$ \\		
			$\nu_0({\rm Hz})$ & $8.7239\times10^{-4}$  &$1.7200\times10^{-8}$  &$10$ \\		
			\hline
			$\delta \tilde{Q}$ &\makecell{$0$ $\qquad \quad $(zero bumpy injection) \\ $3.1\times10^{-3}$ (non-zero bumpy injection)}  &$7.8690\times10^{-4}$  &$20$\\		        		
			\hline 		\hline
			$\theta_s$ &$2\pi/3$  &$1.4977\times10^{-2}$  &$[0, 2\pi]$\\	
			$\phi_s$ &$5\pi/3$  &$2.2273\times10^{-2}$  &$[0, \pi]$\\  
			$\theta_k$ &$\pi/2$  &$1.7072\times10^{-1}$ &$\sim$ \\	
			$\phi_k$ &$0$  &$1.0865\times10^{-1}$ &$\sim$  \\            	        		
			\hline		
		\end{tabular}
		\caption{Injected source parameters and FIM uncertainty estimates for Bumpy-AK signal. The FIM uncertainty estimation is less sensitive to the bumpy parameter so that $\delta \tilde{Q}=0$ is specified to produce the values in the table. Initial phase parameters $(\phi_0$, $\tilde{\gamma}_0$, $\alpha_0)$ are set to zero, with the luminosity distance $D$ scaled to maintain a constant SNR of $50$. To avoid numerical singularities in FIM estimation (observed for $0.5$-year duration signals at SNR $50$), we adopt $1$-year duration signals with SNR $100$. All injected parameters are centered within predefined search ranges to benchmark idealized recovery performance. Investigations of arbitrary parameter placements (e.g., edge cases or multimodal distributions) are deferred to future work.}\label{tab-2-injection}
	\end{table}
	
	Table~\ref{tab-2-injection} lists the typical source parameters adopted from~\cite{Barack:2006pq} with minor adjustments, used to generate the injection signal and their associated parameter uncertainties derived from the FIM. These FIM uncertainties define the baseline search ranges for the analysis. Due to the $7$-dimensional needle-in-a-haystack nature of Bumpy-AK likelihood surfaces, we restrict our search to narrow parameter ranges around the injected values to ensure computational feasibility. While the current work focuses on local parameter estimation, future studies will implement hierarchical search strategies~\cite{Ye:2023lok} to broaden these ranges systematically.
	
	To characterize the algorithm performance in noisy data, we generate $30$ independent noise realizations for each injection in Case $1$ and Case $2$.  The choice of $30$ realizations balances statistical rigor with computational constraints, as detailed in Sec.~\ref{Implementation-cost}.

\begin{figure}[htbp]
		\centering 
	\includegraphics[height=4.0in]{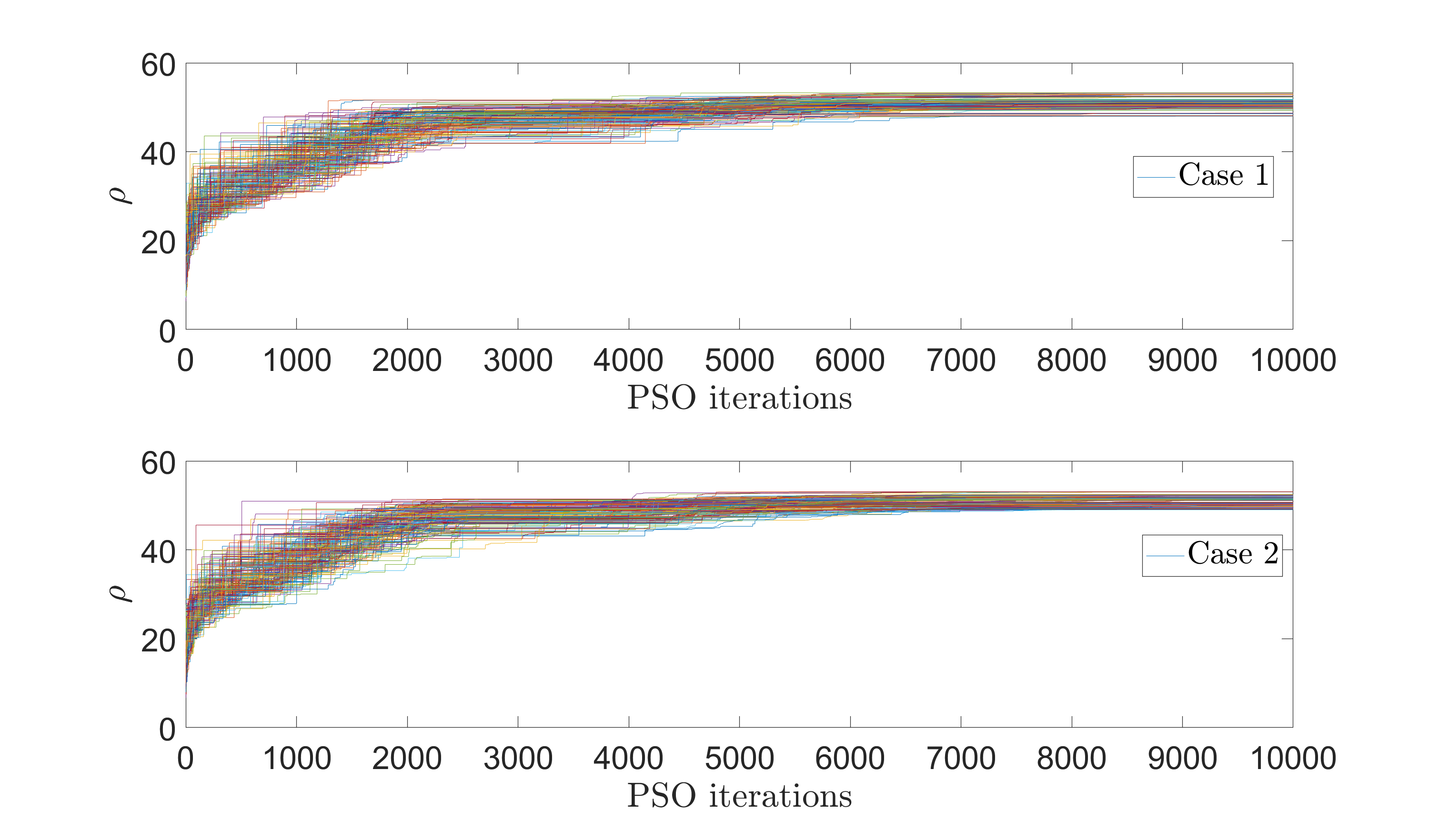}
		\caption{PSO convergence across $180$ independent searches ($30$ noise realizations $\times  6$ trials each), demonstrated by the stabilization of LLR values (Eq.~(\ref{eq:glrt1})) in later iterations. The LLR fluctuations around the injected SNR of $50$ reflect variations induced by distinct noise realizations, with deviations quantifying the interplay between stochastic noise and algorithm performance.}
		\label{fig-pso_convergence}  
\end{figure}
	
\subsection{Case $1$, Zero Bumpy Injection}
\label{Case1}
We generate the injected signal with non-Kerr deviations ($\delta \tilde{Q}=0$) using Eq.~(\ref{eq-hphc}) and Eq.~(\ref{eq:TDI-2}), combining it with $30$ independent noise realizations produced by \texttt{LISACode}~\cite{Petiteau:2008zz} to synthesize $30$ distinct simulated data realizations. For each data realization, we perform $6$ independent PSO searches using a Bumpy-AK template, totaling $180$ searches. As shown in the upper panel of Fig.~\ref{fig-pso_convergence}, all searches achieve robust convergence, evidenced by the stabilization of the LLR values (Eq.~(\ref{eq:glrt1})) in later iterations.
	
\begin{figure}[htbp]
	\centering 
	\includegraphics[height=4.0in]{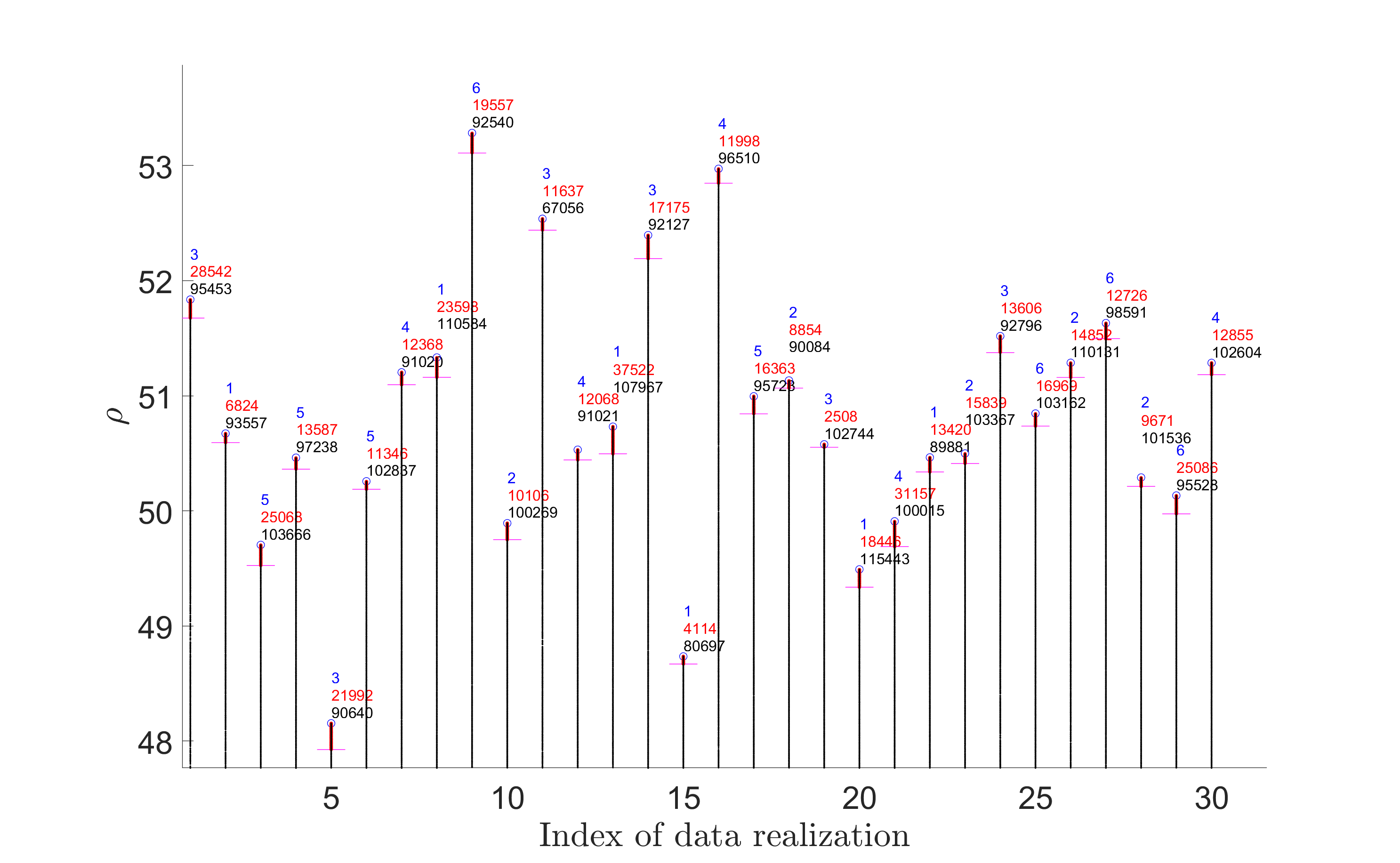}
	\caption{Illustration of the PSO outputs for Case $1$. For each data realization, the $\rho$ (defined in Eq.~(\ref{eq:glrt1})) values at the injection are denoted as a short horizonal line in magenta, the blue circles correspond to the best {\it gbest} point across $6$ independent searches,  the scatter points in red represent the SIPs, while the black denote the STPs where $\rho_{\rm threshold}=20$ is used~\cite{Chua:2021aah}. See Sec.~\ref{SIP-STP} for the definition of SIPs and STPs. The index of the best {\it gbest} point varys from $1$ to $6$, labeled by three rows from top to bottom along with the number of SIPs and STPs, using the color blue, red, and black respectively.}
	\label{Res_c1_1}  
	\end{figure}

\begin{figure}[htbp]
	\centering 
	\includegraphics[height=4.0in]{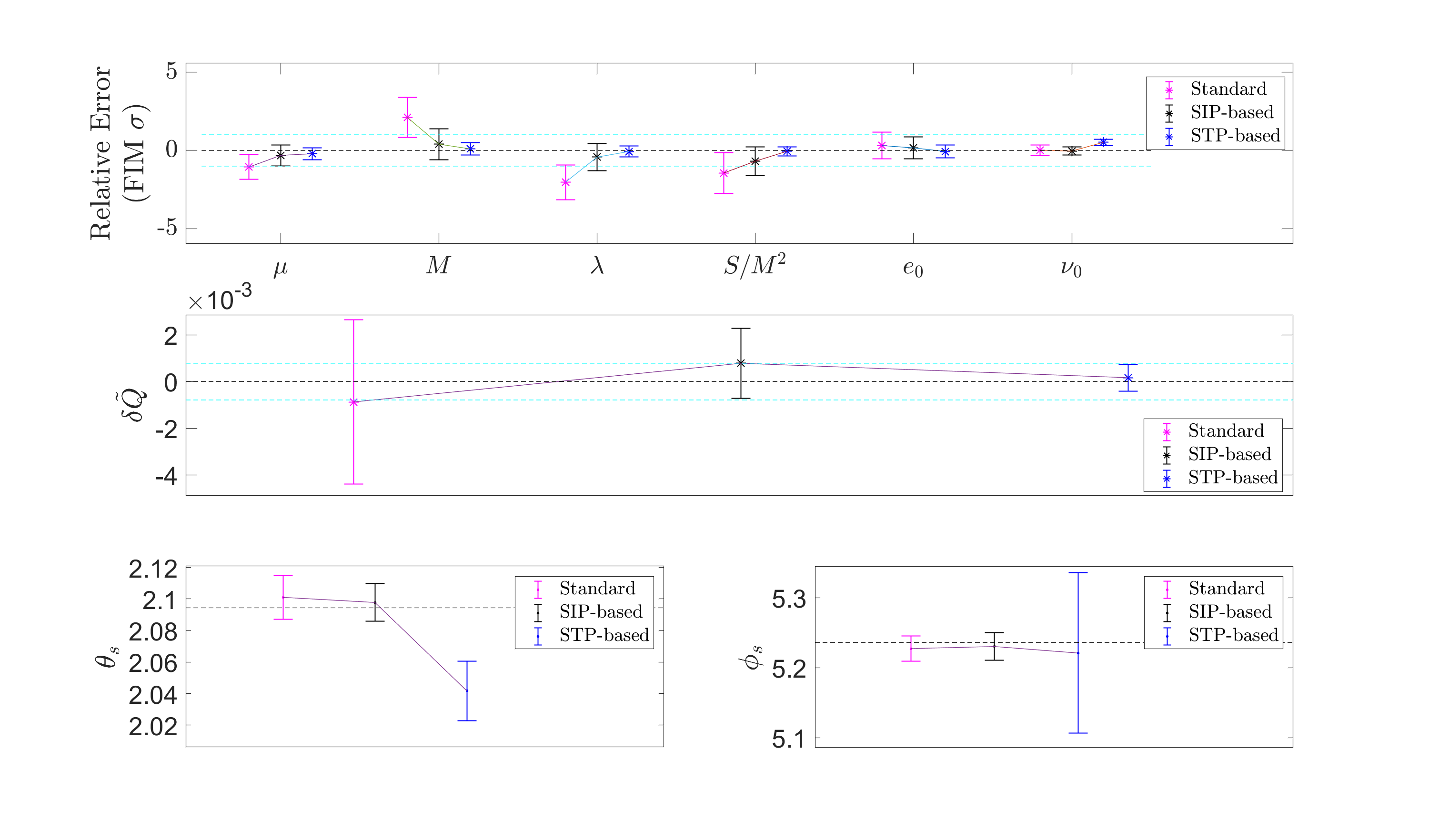}
	\caption{Parameter estimation errors for $\text{C1-1}$, $\text{C1-2}$ and $\text{C1-3}$ in Case $1$. To standardize comparisons across parameters spanning multiple orders of magnitude, the estimation errors for ($\mu$, $M$, $\lambda$, $S/M^2$, $e_0$, $\nu_0$) are calculated as relative deviations from their injected values and normalized by their respective FIM uncertainties, while for ($\delta \tilde{Q}$, $\theta_s$, $\phi_s$) we show the absolute errors. The cyan dashed lines denote $\pm 1 \sigma$ FIM confidence intervals, while the black dashed lines indicate the true injected parameter values. See the definition of the $3$ error bars in the text.}\label{Res_c1_2}
	\end{figure}

Following the standard SNR-threshold value of $20$ corresponding to a detection efficiency of $0.06$~\cite{Chua:2021aah}, the LLR threshold $\rho_{\rm threshold}=20$ is adopted for STPs referring to Sec.~\ref{LLR-threshold}. The counts of SIPs and STPs across each data realization are summarized in Fig.~\ref{Res_c1_1}, demonstrating the algorithm’s capacity to resolve astrophysical signals despite noise fluctuations and parameter degeneracies. 
Fig.~\ref{Res_c1_2} presents the simplest ensemble statistics, sample mean and standard deviation, for the standard, SIP-based, and STP-based parameter estimation defined in Sec.~\ref{SIP-STP}, labeled as $\text{C1-1}$, $\text{C1-2}$, $\text{C1-3}$ respectively.
The three error bars (from $\text{C1-1}$ to $\text{C1-3}$) are aligned horizontally for each parameter in Fig.~\ref{Res_c1_2}, with the following key observations:
	\begin{enumerate}
		\item 
		For the phase-coupled parameters ($\mu$, $M$, $\lambda$, $S/M^2$, $e_0$, $\nu_0$), the $\text{C1-1}$ exhibit significant errors in their fitted means and achieve the largest estimated standard deviations. In contrast, the $\text{C1-2}$ and $\text{C1-3}$ shows reduced errors in their fitted means and smaller standard deviations compared to $\text{C1-1}$, highlighting improved robustness in their parameter estimation by accounting SIPs and STPs.
		\item  For the bumpy parameter  $\delta \tilde{Q}$,  the errors in both its fitted means and stand deviations become smaller from  $\text{C1-1}$ to $\text{C1-2}$ and $\text{C1-3}$ as the involvement of SIPs and STPs, especially for the latter.
		\item 
		For the sky position of ($\theta_s$, $\phi_s$), $\text{C1-1}$ and $\text{C1-2}$ share similar fitted means and standard deviations, while the errors of the fitted means,  and the standard deviations, both increase noticeably for $\text{C1-3}$, reflecting the side-effects of STPs in estimating the sky positions. Note the degeneracy between $(\theta_s,\phi_s)$ and $(\pi-\theta_s,\phi_s+\pi)$ is considered.
	\end{enumerate}  
	
These results demonstrate that incorporating SIPs and STPs significantly improves parameter estimation accuracy compared to relying solely on sparse representative solutions. This enhancement is particularly pronounced for phase-coupled parameters in parameter subset $\theta_1$, where a clear trend emerges as SIPs and STPs are progressively integrated into the analysis, the estimated errors systematically decrease. This underscores the importance of leveraging the full statistical information encoded in these points. However, this benefit comes with trade-off that the accuracy of sky position estimation degrades as the excessive inclusion of STPs. The broader dispersion of STPs in angular coordinates amplifies uncertainties in sky position estimates.

	
	
Fig.~\ref{Res_c1_2} demonstrates that incorporating SIPs and STPs can reduces systematic errors for the parameter estimation of Bumpy-AK signals. Table~\ref{tab-c1-c2-PE} (column $2$, $3$, $4$) quantifies this improvement, listing the parameter errors measured by the fitted means for the three parameter estimates (from $\text{C1-1}$ to $\text{C1-3}$). For the optimal statistical outcomes of $\text{C1-2}$, the uncertainties of the phase-coupled parameters ($\mu$, $M$, $\lambda$, $S/M^2$, $e_0$, $\nu_0$) and the marginal errors of the bumpy deviation $\delta \tilde{Q}$ exhibit  $\sim 1\sigma$ of their FIM predictions, while the sky localization ($\theta_s$, $\phi_s$) achieves arcminute precision, with angular errors of $\sim 0.02$ radians ($\sim1.1$ degree). The parameter estimation errors align with theoretical expectations across all the parameters in $\{\theta_1,\theta_2\}$, validating the robustness of the nested optimization framework.
	
	\begin{table}[htbp]
		\centering
		\begin{tabular}{|c|c|c|c|c|c|c|}
			\hline
			\diagbox{Parameter}{Error}   &$\text{C1-1}$ &$\text{C1-2}$  &$\text{C1-3}$ &$\text{C2-1}$ &$\text{C2-2}$  &$\text{C2-3}$\\
			\hline
			$\mu(M_{\odot})$ &$-1.0615$  &$-0.3140$  &$-0.1065$ &$-0.7902$  &$-0.1358$   &$-0.0282$     \\
			$M(M_{\odot})$ &$2.1152$  &$0.4007$   &$0.1011$  &$1.6613$   &$0.4859$    &$0.0615$  \\
			$\lambda$  &$-2.0295$  &$-0.4239$   &$-0.0985$  &$-1.5755$  &$-0.4347$    &$-0.0424$   \\
			$S/M^2$  &$-1.4445$  &$-0.6852$   &$-0.1554$  &$-1.4786 $   &$-0.2385$    &$-0.0105$  \\
			$e_0$  &$0.3129$   &$0.1633$    &$-0.0109$  &$0.1862$   &$-0.0406$    &$-0.0718$   \\
			$\nu_0$  &$0.0091$  &$-0.0389$   &$0.2236$  &$0.0244$   &$0.0351$    &$0.2348$   \\
			\hline
			$\delta \tilde{Q} (10^{-4})$  &$-8.7215$   &$7.8490$    &$2.4971$  &$2.0896$   &$-2.9873$    &$0.7174$   \\		
			\hline
			$\theta_s$  &$ 0.0067$   &$0.0034$    &$-0.0258$   &$0.0042$   &$0.0017$    &$-0.0264$    \\
			$\phi_s$  &$-0.0087$   &$-0.0057$    &$0.0290$   &$-0.0083$   &$-0.0051$    &$0.027$    \\		
			\hline								
		\end{tabular}
		\caption{The values of parameter estimation errors corresponding to Fig.~\ref{Res_c1_2} for $\text{C1-1}$, $\text{C1-2}$, and $\text{C1-3}$ in Case $1$ (and to Fig.~\ref{Res_c2_2} for $\text{C2-1}$, $\text{C2-2}$, and $\text{C2-3}$ in Case $2$ given in the next section, those values are placed in the same table to make it more compact). Relative errors are reported for the parameters ($\mu$, $M$, $\lambda$, $S/M^2$, $e_0$, $\nu_0$), while absolute errors are used for the bumpy parameter $\delta \tilde{Q}$ and the sky position ($\theta_s$, $\phi_s$). The fitted means incorporate the benefit of SIPs and STPs to mitigate estimated parameter errors, ensuring robust error estimates for each parameters in the table, thus they serve as the basis for computing the tabulated uncertainties.}\label{tab-c1-c2-PE}
	\end{table} 
	
	\subsection{Case $2$, No-Zero Bumpy Injection}
	\label{Case2}
	Having validated the pipeline's baseline performance in Case $1$ (zero bumpy injection with $\delta \tilde{Q}=0$), we now deploy the methodology to Case $2$ to assess its sensitivity to resolvable kerr deviations (non-zero bumpy injection with $\delta \tilde{Q}\ne 0$). As proposed in~\cite{Barack:2006pq}, astrophysically plausible values of $\delta \tilde{Q}$ span from $10^{-4}$ to $10^{-2}$; here, we select a moderate deviation of $3.1 \times 10^{-3}$ to probe resolvable non-Kerr features.
	
\begin{figure}[htbp]
		\centering 
		\includegraphics[height=4.0in]{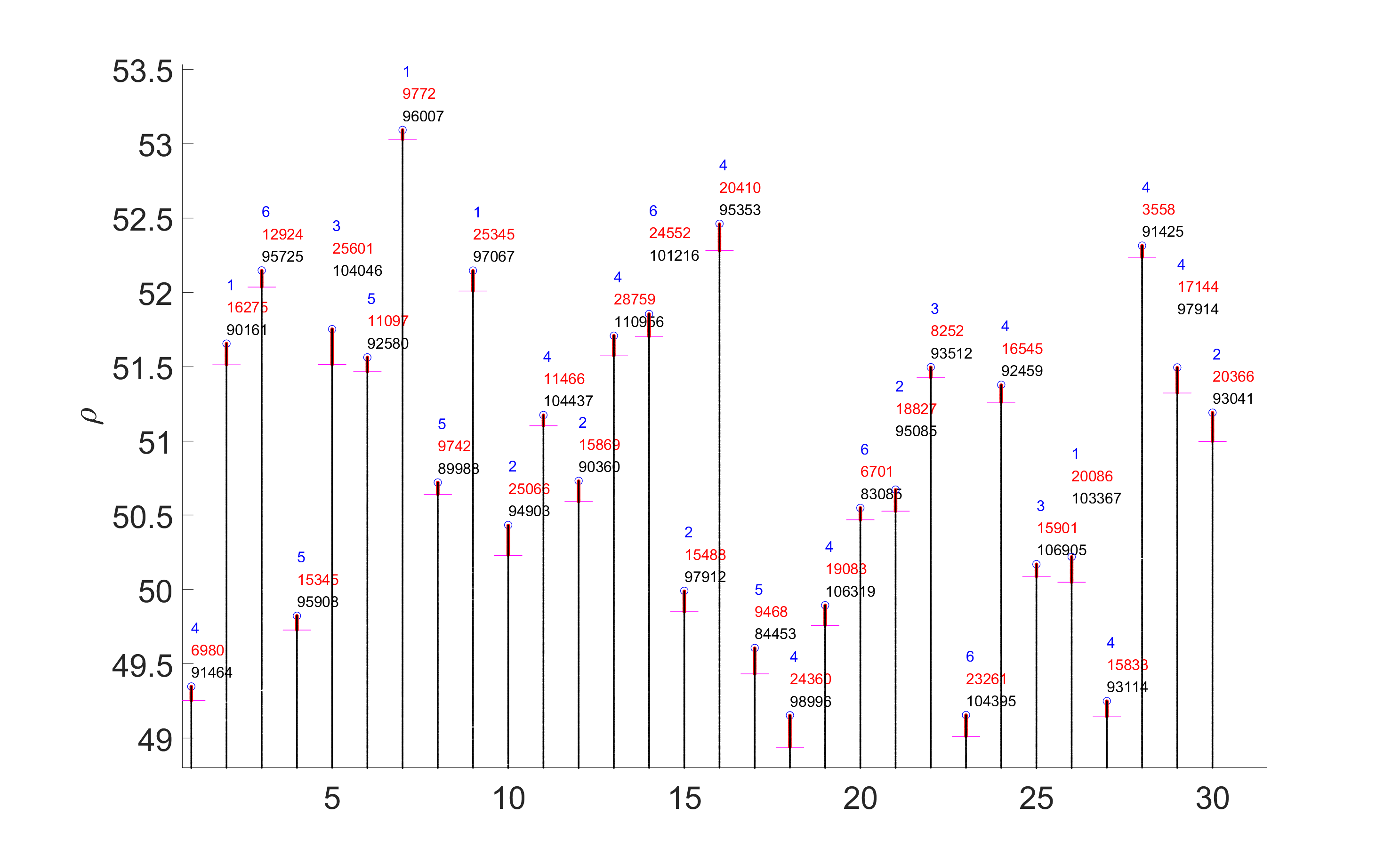}
		\caption{Illustration of the counting of PSO outputs in Case $2$ with the same settings as Case $1$ in Fig.~\ref{Res_c1_1}.}
		\label{Res_c2_1}  
	\end{figure}

\begin{figure}[htbp]
		\centering 
		\includegraphics[height=4.0in]{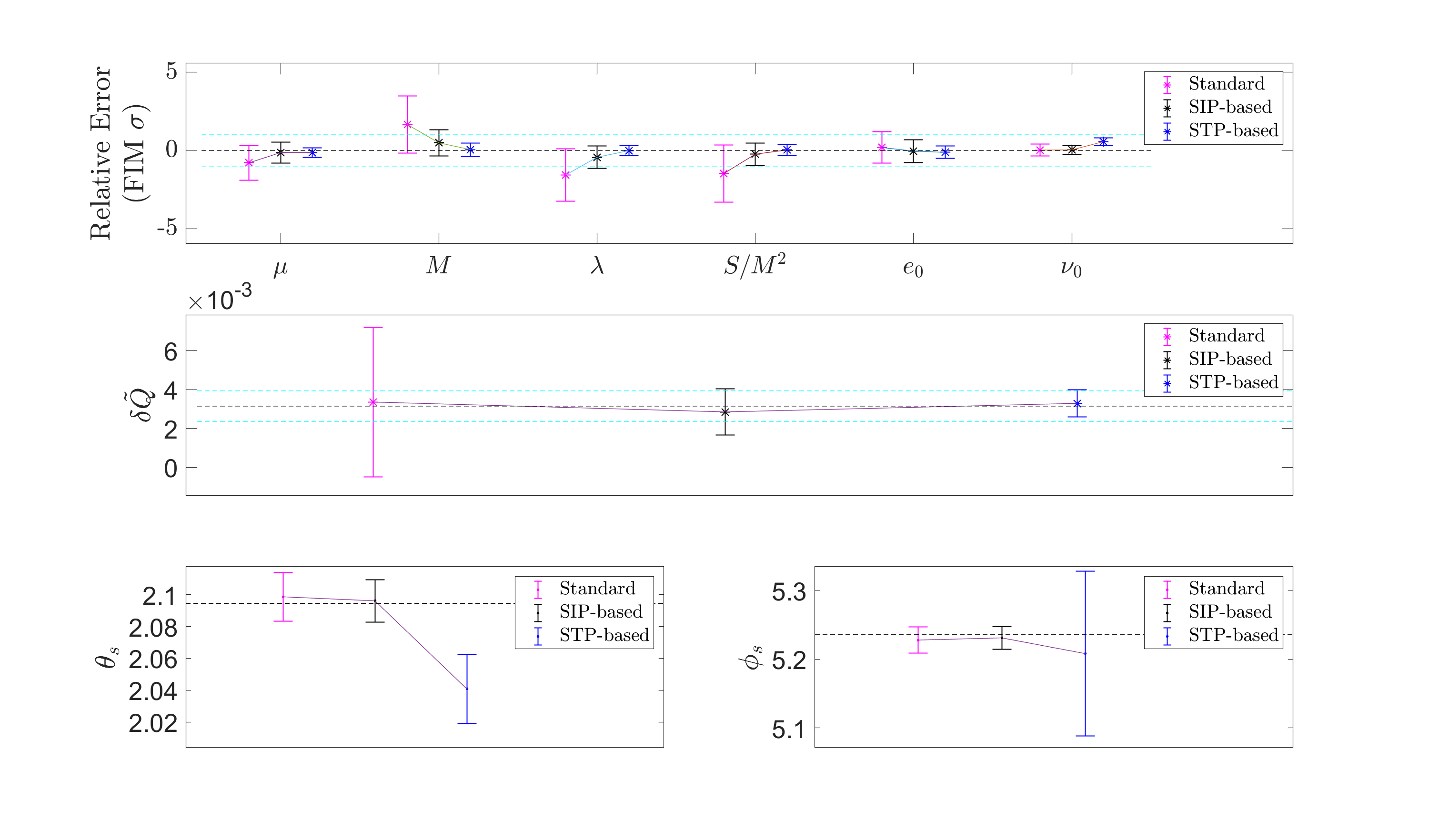}
		\caption{Parameter estimation errors for $\text{C2-1}$, $\text{C2-2}$ and $\text{C2-3}$ in Case $2$ with the same settings as Case $1$ in Fig.~\ref{Res_c1_2}.}
		\label{Res_c2_2}  
	\end{figure}

	Mirroring the workflow of Case $1$, $30$ simulated data realizations are generated by combining the non-zero bumpy signal (using  $\delta \tilde{Q}=3.1 \times 10^{-3}$, Eq.~(\ref{eq-hphc}), and Eq.~(\ref{eq:TDI-2})) with $30$ independent noise realizations produced via \texttt{LISACode}~\cite{Petiteau:2008zz}. For each data realization, $6$ independent PSO searches are executed, totaling $180$ searches. As shown in the lower panel of Fig.~\ref{fig-pso_convergence}, all $180$ searches achieve robust convergence, evidenced by the stabilized LLR values in later iterations. The distribution of PSO outputs  in Fig.~\ref{Res_c2_1} also reveals tens of thousands of SIPs and STPs, consistent with the findings of Case $1$. To ensure detection robustness for unknown injections, we retain the LLR threshold $\rho_{\rm threshold}=20$~\cite{Chua:2021aah} for STPs as discussed in Sec.~\ref{LLR-threshold}.
	Three parameter estimates following the methodology outlined in Sec.~\ref{SIP-STP} are conducted, labeled from $\text{C2-1}$ to $\text{C2-3}$. Each estimates provides fitted means and standard deviations for the parameters in $\{\theta_1,\theta_2\}$, visualized in Fig.~\ref{Res_c2_2}. Key observations mirror those from Case $1$:
	\begin{enumerate}
		\item Incorporating SIPs and STPs achieve significantly errors reduction for the fitted means of phase-coupled parameters in $\theta_1$,  with smaller estimated standard deviations.
		\item The estimated errors of sky localization ($\theta_s$, $\phi_s$) are significantly increased once the STPs are excessively included, otherwise they  remain similar levels. The degeneracy between $(\theta_s,\phi_s)$ and $(\pi-\theta_s,\phi_s+\pi)$ is also considered.
	\end{enumerate}
	These parallels underscore the robustness of the nested optimization framework for non-zero bumpy signal recovery.
	
	Table~\ref{tab-c1-c2-PE} (column $4$, $5$, $6$) lists the parameter estimation errors for Case $2$, including the fitted means of parameters in $\{\theta_1, \theta_2\}$ for all the estimates ($\text{C2-1}$ to $\text{C2-3}$). For the optimal estimate of $\text{C2-2}$, the estimated parameter errors align with theoretical FIM expectations: the phase-coupled parameters ($\mu$, $M$, $\lambda$, $S/M^2$, $e_0$, $\nu_0$) exhibit uncertainties within $1 \sigma$ of their FIM length and the bumpy parameter $\delta \tilde{Q}$ reach the uncertainties within $\%10$ of the injected non-zero value. While sky localization angles ($\theta_s$, $\phi_s$) achieve arcminute precision with angular errors of $0.02$ radians ($\sim 1.1)$ degree. These results validate the framework’s robustness in resolving non-Kerr signatures under moderate deviations $(\delta \tilde{Q}=3.1\times10^{-3})$.

	\subsection{Exploring the distribution of SIPs}
	Our analysis reveals that the one-dimensional slice of the bumpy parameter $\delta \tilde{Q}$ for SIPs exhibit broad distributions in both Case $1$ and Case $2$, with many solutions approaching the boundaries of the prescribed search range. To better characterize this distribution, we conduct additional searches using an expanded search range for $\delta \tilde{Q}$ while maintaining fixed search ranges for the other parameters (Table~\ref{tab-2-injection}). Due to computational constraints, this investigation is limited under single noise realization from \texttt{LISACode}~\cite{Petiteau:2008zz}, analyzed through $6$ independent PSO searches. We employ two distinct injection configurations with varying $\delta \tilde{Q}$ values and different search ranges to systematically examine boundary effects in its parameter recovery for both Case $1$ and Case $2$.
	\begin{enumerate}
		\item 
		Fig.~\ref{Res_c12_1} displays the parameter estimation outcomes with the injection centered in the search space while progressively increasing the width of $\delta \tilde{Q}$ from $30\sigma$ to $40 \sigma$,  $100 \sigma$, respectively, where $\sigma$ denotes its FIM uncertainty (Table~\ref{tab-2-injection}). 
		The best-fit values for $\delta \tilde{Q}$ are distributed within the predicted range of $[-0.01,0.01]$, consistent with theoretical expectations from~\cite{Barack:2006pq}. Analysis of pairwise distances between those solutions reveals no instances of overlapping points, demonstrating that the broad yet densely distributed SIPs present in the parameter space.
		
		\begin{figure}[htbp]
			\centering 
	\includegraphics[height=4.0in]{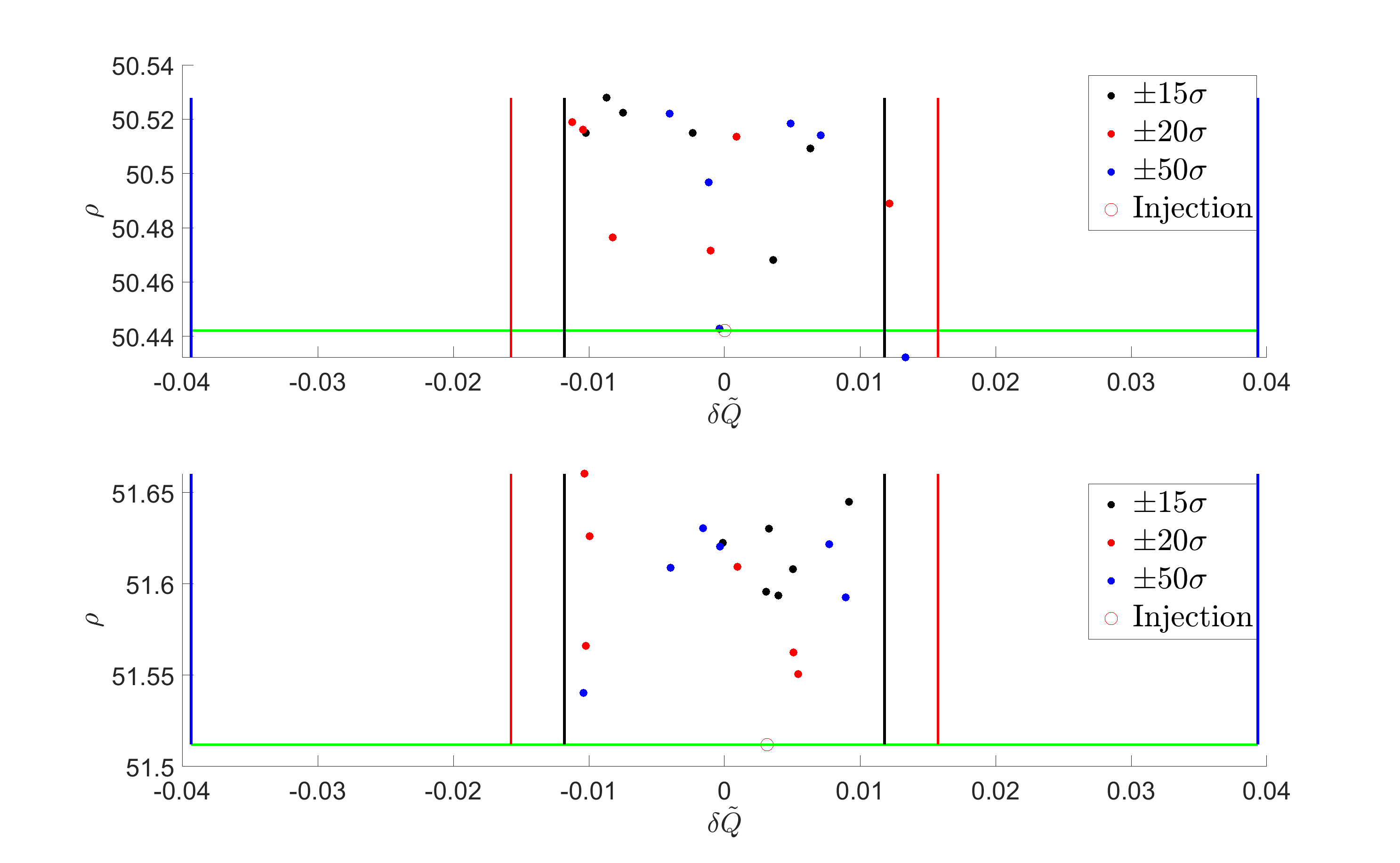}
			\caption{Illustration of the PSO outcomes with inside injection. The upper panel corresponds to Case $1$ ($\delta \tilde{Q}=0$), and the bottom panel shows Case $2$ ($\delta \tilde{Q}=3.1\times 10^{-3}$). The best-fit parameters derived via PSO are represented as scatter points in black, red, and blue, respectively, with their associated search ranges $(30\sigma$, $40 \sigma$,  $100 \sigma)$ demarcated by vertical lines in corresponding colors, while the injection is demarcated by a red circle and the LLR reference value at the injection in terms of $\rho$ defined in Eq.~(\ref{eq:glrt1}) is plotted as a horizontal solid green line. The $\sigma$ of $\delta \tilde{Q}$ is $7.8690\times10^{-4}$ given in Tab.~\ref{tab-2-injection}.}
			\label{Res_c12_1}  
		\end{figure}	
		\item
		Fig.~\ref{Res_c12_2} demonstrates the parameter estimation outcomes of $\delta \tilde{Q}$ when its injection lies outside the search range, which employs a range width of $5 \sigma$, and spans $5 \sigma$ in both directions from the injection value with $\sigma$ denoting its FIM uncertainty (Table~\ref{tab-2-injection}).  Remarkably, SIPs persist despite the injection being excluded from the $\delta \tilde{Q}$ search range. This observation reveals that the involvement of the bumpy parameter result in an extremely multimodal structure that remains largely insensitive to the injection location, even when examining regions far removed from the true parameter values.
		
		\begin{figure}[htbp]
			\centering 
	\includegraphics[height=4.0in]{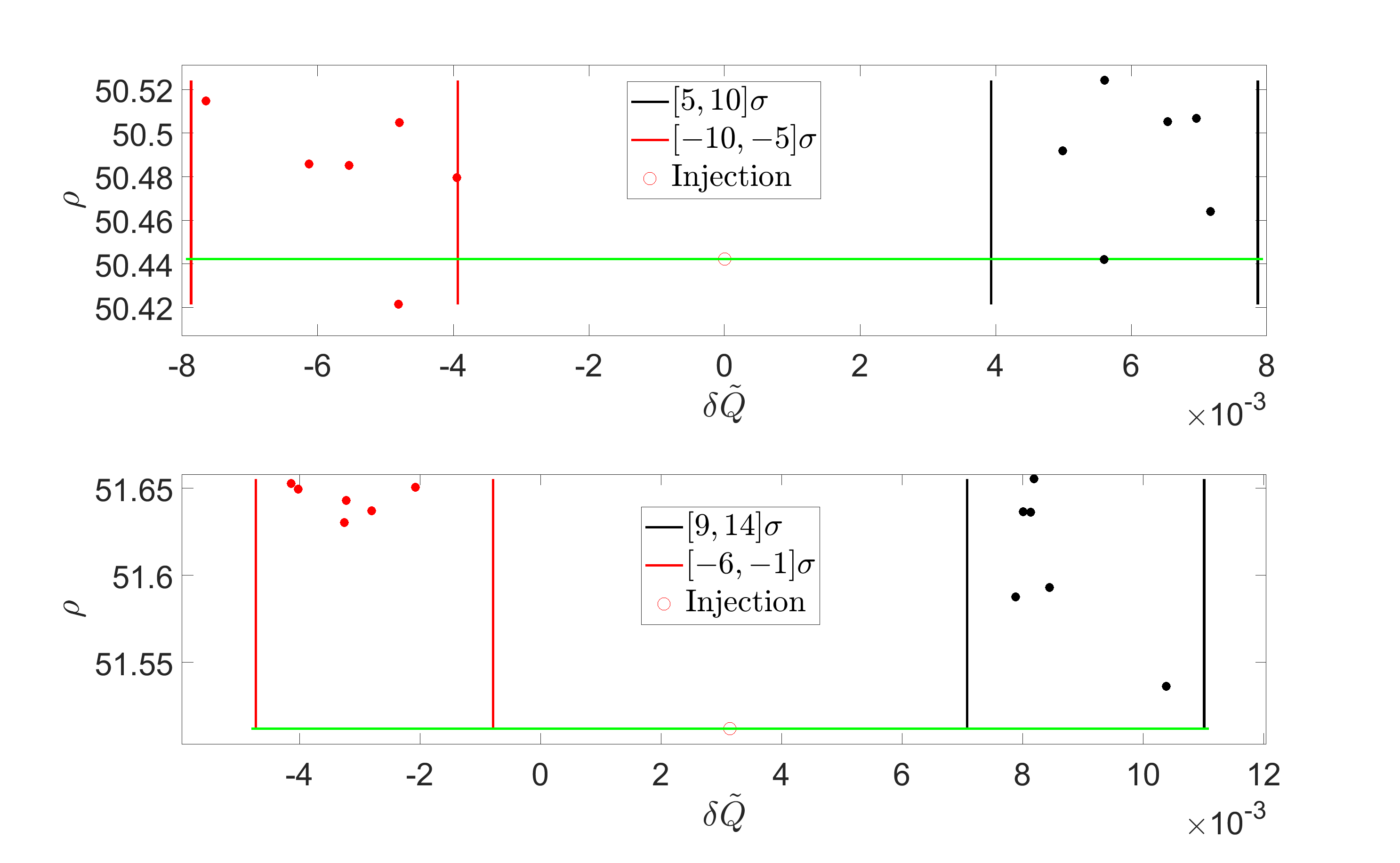}
			\caption{Same with Fig.~\ref{Res_c12_1} but for outside injection of $\delta \tilde{Q}$. The $6$ best-fit solutions for each search and their corresponding search ranges are color-coded, with the red and black markers representing left-side and right-side searches respectively.}
			\label{Res_c12_2}  
		\end{figure}	
	\end{enumerate}  
	
	The existence of SIPs under these conditions highlights the complex morphology of the likelihood surface for Bumpy-AK waveform. Therefore more attentions are deserved to develop the strategy dealing with those SIPs in extending the workhorse targeting the waveform from AK to Bumpy-AK. In future advanced EMRI data analysis, the AK would be replaced with more accurate waveform, e.g., Fast EMRI Waveforms (FEW)~\cite{Katz:2021yft}. Similar SIP issues are expected to occur and the corresponding strategy are indispensable for FEW to constraint non-kerr features.
	
	\subsection{Tuning of the LLR Threshold $\rho_{\rm threshold}$}
	\label{LLR-threshold}
	The GW detection is statistically significant when the best-fitted square root of LLR value exceeds a predetermined threshold, e.g.,  $\rho_{\rm threshold}$ in Eq.~(\ref{eq:SIP-STP}). In practical applications, this threshold for candidate events is calibrated against a false alarm rate (FAR) using a large number of noise realizations. As the limited computational resources in hand,  we relegate such a rigorously computed threshold to future work once our code is Graphic Processing Unit (GPU) enabled in the next version. A simple theoretical estimation between FAR and $\rho_{\rm threshold}$ is given in~\cite{Chua:2017ujo} where $\rho_{\rm threshold}=-2\log({\rm FAR})$. For the threshold value $\rho_{\rm threshold}=20$ adopted in section~\ref{Case1} and ~\ref{Case2}, the corresponding FAR is as low as $4.54\times10^{-5}$, indicating that noise-induced $\rho$ values have a negligible probability of exceeding this threshold.
	We correlate the Bumpy-AK waveform template with a single pure noise realization generated via \texttt{LISACode}~\cite{Petiteau:2008zz} for a direct and insufficient check. $6$ independent PSO searches are conducted using different random seeds for generating initial positions and velocities, with the resulting maximum $\rho$ value (defined in Eq.~(\ref{eq:glrt1})), summarized in the second column of Table~\ref{tab-3-LLR-threshold}. The resulting $\rho$  values are significantly lower than the predefined threshold $\rho_{\rm threshold}=20$. Consequently, any $\rho$ value exceeding this threshold can be confidently attributed to GWs.
	\begin{table}[htb]
		\centering
		\begin{tabular}{|c|c|c|c|c|c|}
			\hline
			\diagbox{PSO}{Case}   &Case 1 and Case 2 &\makecell{Case 3 \\SNR=$20$} &\makecell{Case 3 \\SNR=$50$} &\makecell{Case 3 \\SNR=$100$} &Case 4\\
			\hline
			$1$ &$7.099570$ &$16.0232$ &$29.8263$ &$57.6315$ &$7.1843$ \\
			$2$ &$7.382015$ &$15.8366$ &$29.7987$ &$57.6337$ &$7.0259$ \\
			$3$ &$6.891719$  &$15.7858$ &$29.7106$ &$57.4129$ &$7.0194$ \\
			$4$ &$6.752540$ &$15.8685$ &$29.8668$ &$57.5970$ &$7.2680$ \\
			$5$ &$6.988922$ &$15.7914$ &$29.8016$ &$57.7375$ &$6.9523$ \\
			$6$ &$6.924624$  &$16.0438$ &$29.7113$ &$57.3857$ &$7.4978$ \\
			\hline
		\end{tabular}
		\caption{The $\rho$ values at {\it gbest} returned by each PSO search in different cases, placed together to make the table more compact. The column $1$ enumerates the index of the $6$ independent PSO searches performed for each scenario. The column $2$ corresponds to checking the choice of $\rho_{\rm threshold}=20$ by matching Bumpy-AK templates waveform with a pure noise. Columns $3,4,5$ present results for Case $3$ (small waveform inaccuracy tests) at SNRs of $25$, $50$, and $100$, respectively. The column $6$ reports $\rho$ at {\it gbest} for Case $4$, which investigates larger waveform mismatches.}\label{tab-3-LLR-threshold}
	\end{table} 
	
	\begin{figure}[htbp]
		\centering 
		\includegraphics[height=4.0in]{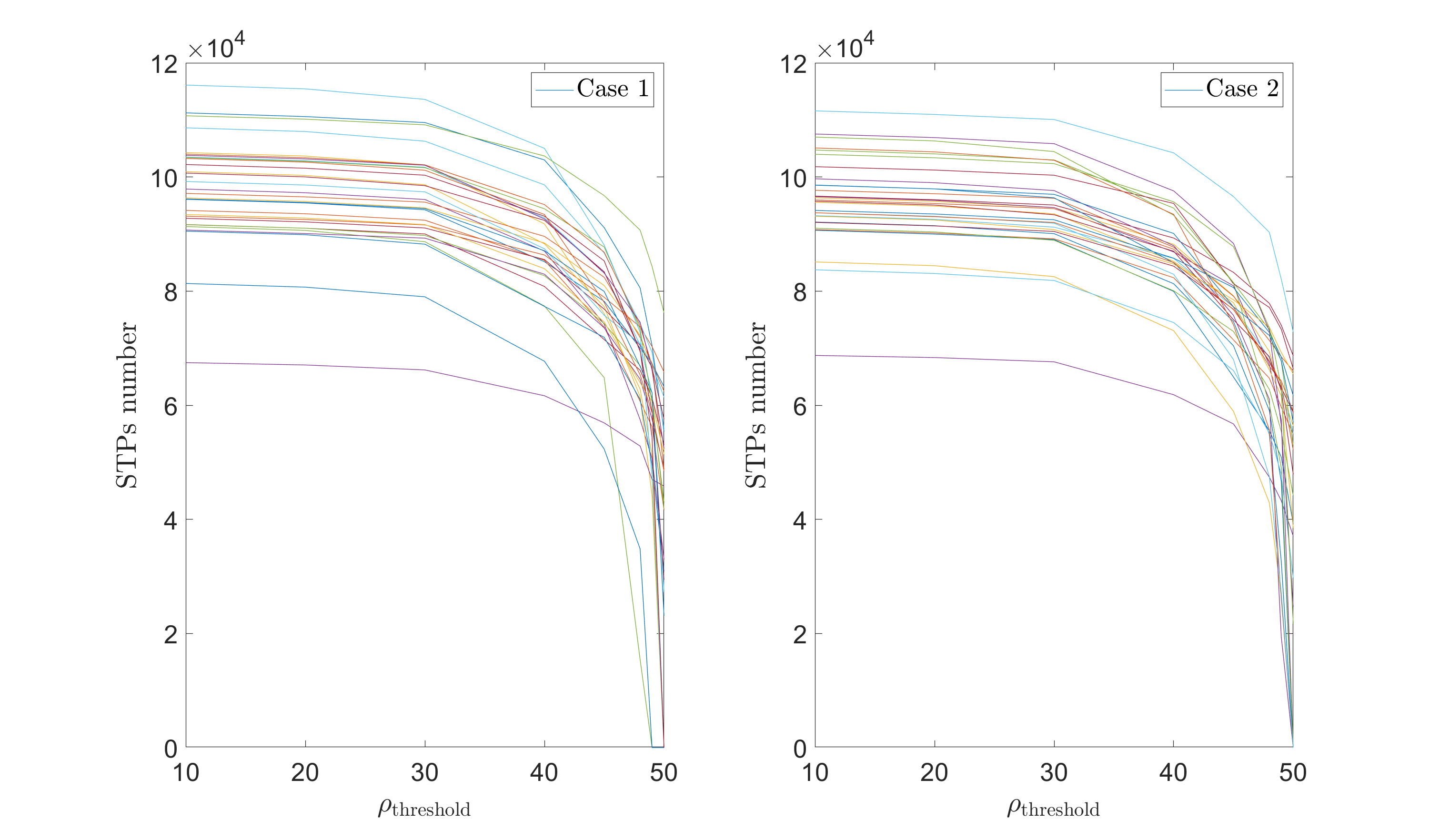}
		\caption{Illustration of how the STPs number scale with the $\rho_{\rm threshold}$ for each data realization (total $30$) in both Case $1$ (left panel) and Case $2$ (right panel) using narrow prior in Table~\ref{tab-2-injection}.}
		\label{fig:STP-threshold}  
	\end{figure}
	The selection of alternative $\rho_{\rm threshold}$ is also investigated. As illustrated in Fig.~\ref{fig:STP-threshold} for both Case $1$ and Case $2$, the number of STPs exhibits minimal variation when $\rho_{\rm threshold}$ lies within the range of $[10,30]$. This stability arises from the use of a narrow prior distribution (Table~\ref{tab-2-injection}), which enables the PSO algorithm to efficiently traverse low-$\rho$ regions ($\rho < \sim 30$) within the parameter space using relatively few iterations. A consistent trend is observed in Fig.~\ref{fig-pso_convergence}, where PSO achieves convergence in approximately $5000$ iterations, despite a total budget of $10,000$ iterations. Consequently, tuning $\rho_{\rm threshold}$ ​in this study is straightforward, and any value within $[10,30]$ may be selected without significantly affecting the results.
	\subsection{Case $3$, Small template uncertainty}
	Accurate modeling of EMRI waveforms remains a significant challenge in GW astronomy. The approximate kluge waveform has gained widespread use in current EMRI data analyses due to its computational efficiency, despite its substantial deviations from true waveforms, particularly when self-force effects are considered. This inherent template uncertainty will inevitably affect future EMRI data analysis. In our investigation, we focus on small-scale uncertainties by employing the standard AK model to generate injected signals using the same parameters given in Tab.~\ref{tab-2-injection}. The AK injection is derived from the orbital ODEs in Eq.~(\ref{eq:ODEs}) while deliberately excluding the $\tilde{Q}$ terms, inducing controlled waveform inaccuracies. Three SNRs injections (($25$, $50$, $100$) by rescaling the distance $D$ of the AK signal with other parameters fixed as Tab.~\ref{tab-2-injection}) are used to examine the SNR dependence of parameter estimation under small template uncertainties. The interesting question is whether such waveform uncertainties could result in fake bumpy predictions inspired by ref~\cite{Shen:2025svs}. 
	
	Due to computational constraints, a single noise realization produced by \texttt{LISACode}~\cite{Petiteau:2008zz} is combined with the three injected AK signals, leading to three data realizations. We conduct matched filtering between the Bumpy-AK template and each data realization using $6$ independent PSO searches. Following the methodology established in Case $1$ and Case $2$, we focus on STPs to simulate a more realistic scenario where the injection parameters are unknown. Applying the threshold of $\rho_{\rm threshold}=20$ refering to Sec.~\ref{LLR-threshold}, we identify valid STPs for two different injected SNRs values ($50$, $100$) whose accumulated counts across all $6$ searches are $84998$ and $89133$ respectively. As for the injection of SNRs value $25$, the threshold of $\rho_{\rm threshold}=10$ corresponding to a higher FAR is used with $87064$ STPs because it is none for the threshold of $\rho_{\rm threshold}=20$.  
	
\begin{figure}[htbp]
		\centering 
	    \includegraphics[height=4.0in]{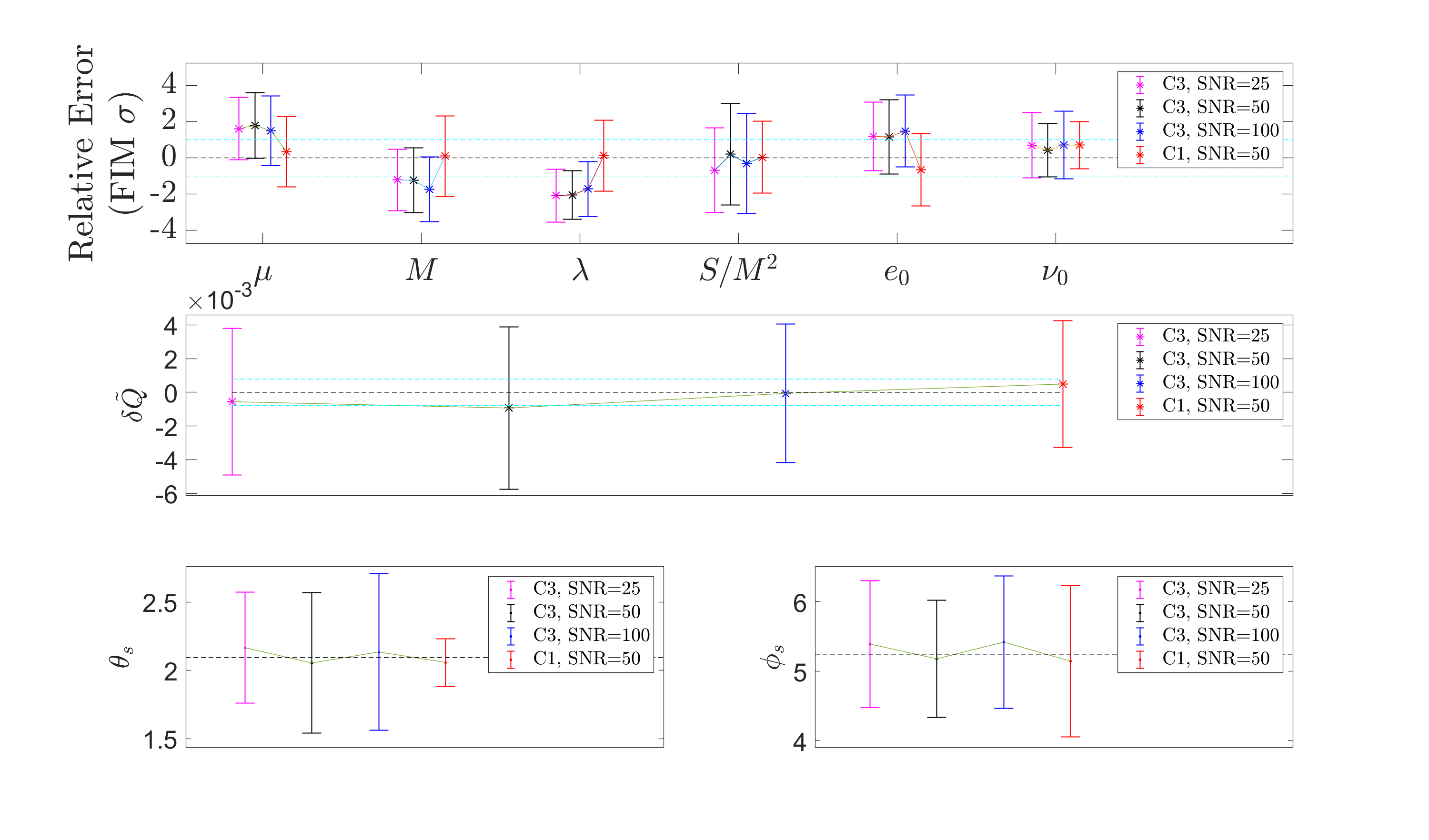}
		\caption{Illustration of the parameter estimation errors for Case $3$.  Each parameter displays four error bars: three (magenta, black, blue) correspond to SNR=$\{25,50,100\}$ in Case $3$, ordered left to right, while the $4$th in red represents STPs identified in Case $1$ under the identical noise realization and threshold $\rho_{\rm threshold}=20$, excluding template uncertainty in the injection model. The legend C$1$ and C$3$ correspond to Case $1$ and Case $3$ respectively.}
		\label{Res_c3_1}  
	\end{figure}

	Our analysis reveals significant SNR loss for all the three SNR injections in Case $3$, where small template uncertainties are present, as shown in columns $4,5,6$ of Table~\ref{tab-3-LLR-threshold}. 
	The true LLR values exhibit minor fluctuations around the nominal SNR injection due to noise effects, though they remain closely correlated with the injected SNR levels ($25$, $50$, $100$), respectively. 
	Fig.~\ref{Res_c3_1} presents the parameter estimation in terms of the sample means and standard deviations of all collected STPs across $6$ searches for the SNR ($25$, $50$, $100$), respectively, and the STPs identified in Case $1$ under the identical threshold value $\rho_{\rm threshold}=20$, noise realization and injected SNR of $50$. This comparison demonstrates that
	\begin{enumerate}
		\item 
		For the $6$ phase-coupled parameters shown in the $1$st row of Fig.~\ref{Res_c3_1}, the first $3$ error bars (corresponding to analyses using the less accurate template) reveal significant errors in term of the fitted means and the $1\sigma$ confidence level specified by the standard deviations. The comparison between the $2$nd and $4$th error bars definitively excludes noise effects as the source of these parameter estimation errors, demonstrating that even small template uncertainties inevitably introduce systematic errors in parameter estimation. Furthermore, the observed SNR dependence suggests that these template-induced errors persist across all tested SNR levels below $100$, indicating that such systematic errors remain unavoidable within this SNR range regardless of signal strength.
		
		\item  For the bumpy parameter $\delta \tilde{Q}$, analysis reveals weak evidence that higher SNR marginally reduce errors in fitted means when using the less accurate template. The SNR $100$ case (Case $3$) exhibits the smallest errors across tested SNRs ($25$, $50$, $100$), outperforming even SNR $50$ of Case $1$ configuration where template uncertainty was excluded, demonstrating how higher SNR improves parameter estimation fidelity of  $\delta \tilde{Q}$ despite small inherent waveform modeling uncertainties. However, the larger errors observed for SNR $50$ relative to SNR $25$ (Case $3$) may arise from statistical fluctuations. Notably, the large standard deviations across all cases reflect the broad distribution of STPs in the parameter space.
		
		\item Regarding the sky localization, those fitted means exhibit errors of approximately $0.1$ radians, with standard deviations of similar magnitude. 
	\end{enumerate}
	
	The Case $3$ tests reveal that even minor waveform template uncertainties induce significant errors in phase-coupled parameter estimation (e.g.,$\mu$, $M$, $\lambda$, $S/M^2$, $e_0$, $\nu_0$) in terms of their fitted means. However, the bumpy parameter $\delta \tilde{Q}$ remains robust: its fitted means exhibit unbiased  ($\le 1 \sigma$ deviation from its truth assuming zero, in terms of its FIM uncertainties) despite larger estimated standard deviations, reflecting a bias-variance trade-off inherent to template mismatches. These results suggest limited risk of spurious $\delta \tilde{Q}$ detections under small template errors. Nevertheless, developing more accurate waveform models remains critical to minimize the overall systematic errors for all the phase-coupled parameters in precision GW astronomy.
	\subsection{Case $4$, Large template uncertainty}
	Case $3$ investigates the parameter estimation under small template uncertainties for the Bumpy-AK signal. In contrast, Case $4$ explores large template uncertainties by replacing the AK model with FEW~\cite{Katz:2021yft}, where the FEW parameter $p_0$
	is empirically mapped to the Bumpy-AK parameter $\nu_0$ to establish a physically consistent search range for $\nu_0$. We generate simulated data by combining the FEW injection with single noise realization from \texttt{LISACode}~\cite{Petiteau:2008zz}. $6$ independent PSO searches are conducted, with the resulting $\rho$ (defined in Eq.~(\ref{eq:glrt1})) values of each {\it gbest} tabulated in column $7$ of Table~\ref{tab-3-LLR-threshold}. This methodology enables systematic comparison of template uncertainty impacts across waveform families while maintaining computational feasibility.

    The best-fit $\rho$ values obtained through parameter estimation (column $7$ of Table~\ref{tab-3-LLR-threshold}) are notably lower than the present detection threshold $\rho_{\rm threshold}=20$ and closely align with the noise-induced $\rho$ levels (column $2$ of Table~\ref{tab-3-LLR-threshold}). This proximity suggests considerable challenges in recovering reliable signals under significant template waveform uncertainties, which may lead to missed detections in practical applications. Furthermore, this underscores a fundamental limitation of the current Bumpy-AK template, it appears insufficiently accurate to detect and characterize signals generated by more realistic and complex waveform models, such as those produced by FEW.
	\subsection{Extension to a true search algorithm}
	Previous findings have demonstrated successful local parameter recovery in the vicinity of the true solution when using narrow priors; however, this does not constitute a global detection and characterization algorithm. A truly robust search algorithm must operate effectively across wide priors for all phase-coupled parameters in the EMRI waveform. Nevertheless, the needle-in-a-haystack nature of EMRI signals has historically impeded the use of wide priors in data analysis, until the recent development of hierarchical search algorithms referenced in~\cite{Ye:2023lok,Strub:2025dfs,Cole:2025sqo}. These methods iteratively narrow the prior ranges and overcome the non-local degeneracies inherent in EMRI parameter estimation~\cite{Chua:2021aah}. In this subsection, we conduct a proof-of-concept hierarchical search for at least one parameter under a realistically broad prior to demonstrate the feasibility of extending this approach toward a complete search algorithm. We focus specifically on the massive black hole mass $M$ and the initial orbital frequency $\nu_0$, employing wide priors of $[5\times10^5, 5\times10^6]$ for $M$ and $[10^{-4}, 0.1]$ Hz for $\nu_0$​, while the remaining six phase-coupled parameters (including $\delta \tilde{Q}$​) retain the narrow priors specified in Table~\ref{tab-2-injection}.
	
\begin{figure}[htbp]
		\centering 
		\includegraphics[height=4.0in]{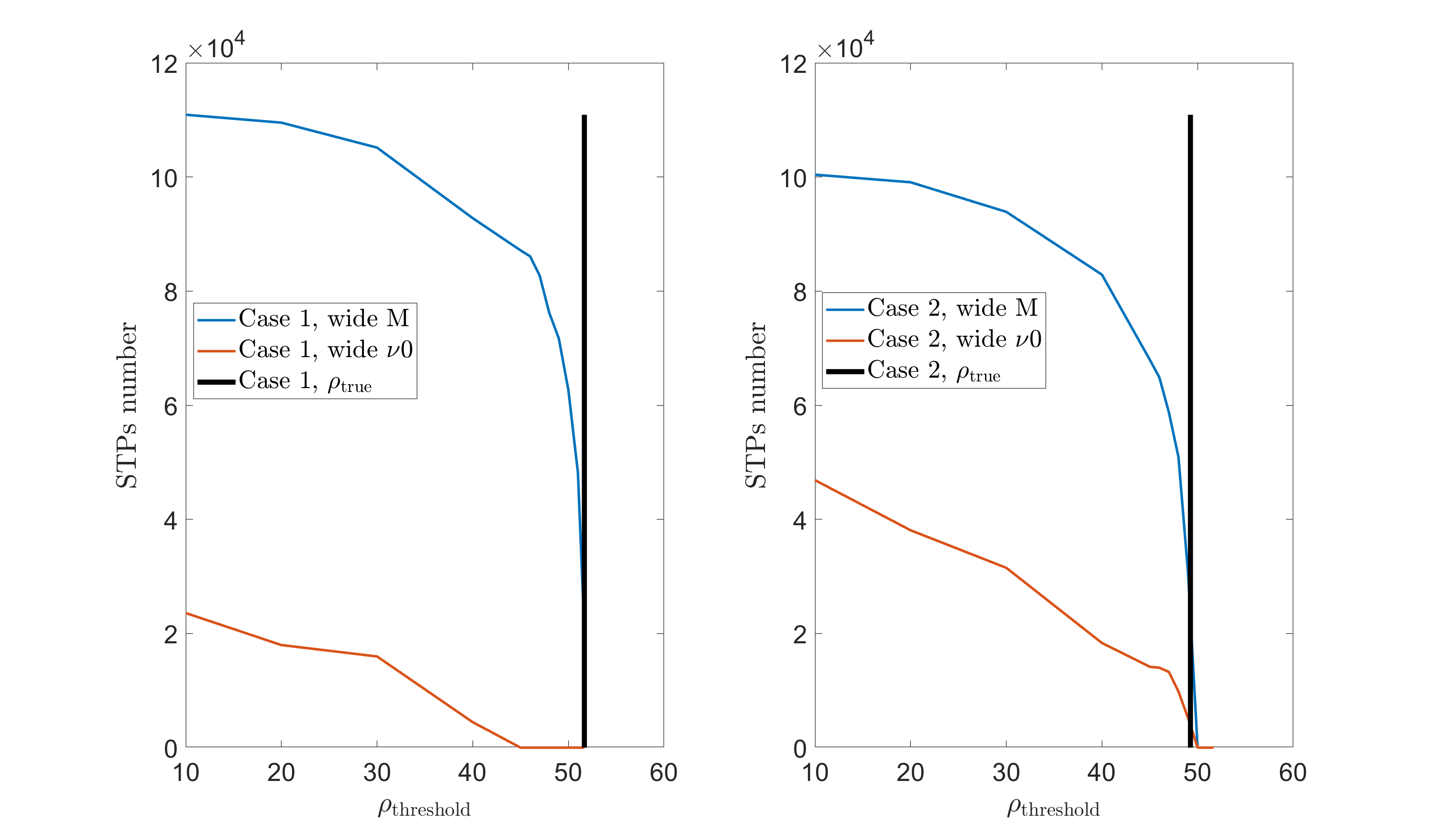}
		\caption{Illustration of how the STPs number scale with the $\rho_{\rm threshold}$ for single data realization (the $1$st among $30$) in both Case $1$ (left panel) and Case $2$ (right panel) using wide prior for $M$ and $\nu_0$, respectively.}
		\label{fig:STP-threshold-wide}  
	\end{figure}

	The resulting number of STPs varies with the choice of $\rho_{\rm threshold}$​, as illustrated in Fig.\ref{fig:STP-threshold-wide}. Two key observations emerge:
		\begin{enumerate}
		\item When applying a wide prior solely to $M$, the number of STPs remains at approximately $10^5$, similar with the narrow prior case shown in Fig.~\ref{fig:STP-threshold} for $\rho_{\rm threshold} \in [10,30]$. Moreover, the PSO algorithm successfully locates degenerate peaks with amplitudes comparable to $\rho_{\rm true}$ for both the Case $1$ and Case $2$.
		\item In contrast, when a wide prior is applied only to $\nu_0$, the number of STPs decreases significantly by approximately an order of magnitude relative to the narrow prior baseline. Although PSO identifies degenerate peaks reaching $\rho_{\rm true}$ in Case $2$, it fails to do so in Case $1$. 
	    \end{enumerate}
	    
	 These results highlight differing coupling strengths between $M$ or $\nu_0$ ​and the remaining phase-coupled parameters. Extending wide priors to all phase-coupled parameters would likely introduce two challenges: (i) heightened sensitivity of the STP count to the specific value of $\rho_{\rm threshold}$, necessitating more careful tuning, and (ii) a substantial reduction in the number of STPs, undermining the reliability of ensemble statistics needed to mitigate parameter estimation errors arising from the inclusion of the bumpy parameter $\delta \tilde{Q}$ ​in the Bumpy-AK waveform.
	 
	 In summary, the simultaneous presence of extra degeneracies between EMRI parameters and the bumpy parameter $\delta \tilde{Q}$​, combined with non-local degeneracies among the EMRI parameters themselves~\cite{Chua:2021aah}
	 , demands a hierarchical search algorithm capable of overcoming this dual-layer degeneracy. Addressing these challenges comprehensively exceeds the scope of this paper and is deferred to future work.   
	\section{Discussion}
	\label{Discussion}
	The GW signals from EMRIs exhibit waveform morphologies characterized by sharp, localized peaks in phase-coupled parameter subspace. Standard EMRI signals inhabit a $6$-dimensional peak, while their ``bumpy" counterparts (incorporating non-Kerr spacetime deviations) extend this to $7$ dimensions. Such high-dimensionality of the full parameter space, coupled with the exquisite sensitivity required for GW phase reconstruction, transforms EMRI data analysis into a challenging needle-in-a-haystack problem. This represents one of the most computationally demanding aspects of space-borne GW detection missions targeting EMRI sources, particularly given the signals' weak amplitudes and long-duration waveforms embedded in overlapping confusion noise.

	The inclusion of the bumpy parameter $\delta \tilde{Q}$ in Bumpy-AK waveform model fundamentally alters the parameter space morphology compared to standard EMRIs, introducing widely distributed SIPs across the range $[-0.01,0.01]$. Beyond the inherent difficulty of a $7$-dimensional sharp peak, robust analysis of Bumpy-AK signals require systematic handling of these SIPs to mitigate parameter estimation errors. This study demonstrates that PSO, through multiple independent searches, efficiently accumulates SIPs. Subsequent statistical aggregation of these points via sample means significantly reduces systematic bias for the phase-coupled parameters in $\theta_1$, while the sky localization maintains acceptable errors ($\sim 0.1$ radians). The broad dispersion of SIPs introduces elevated standard deviations, a trade-off inherent to this methodology. These results highlight the critical role of degeneracy mapping in resolving high-dimensional waveform features while quantifying the precision-recall balance in the parameter estimation of Bumpy-AK signals.
    Note that STPs should be used instead of SIPs  in real-data analyses where its thresholds require precise calibration to optimally approximate the LLR value at the true signal’s embedded location within the observed dataset, thereby reducing the estimation error in sky localization.
	
	In this study, we employed narrow, injection-informed priors and a waveform duration of $0.5$-year to address the "needle-in-a-haystack" challenge inherent in EMRI data analysis. This approach helps control the severity of intrinsic non-local degeneracies among EMRI parameters, thereby isolating the root cause of widely distributed SIPs and STPs. We ensure that these features arise specifically from the inclusion of the bumpy parameter in the Bumpy-AK waveform model, where ensemble averaging over SIPs or STPs remains valid and effectively reduces parameter estimation errors. However, for a robust global detection and characterization algorithm, it is essential to adopt realistically broad priors encompassing all phase-coupled parameters and the signal duration matches the operational lifespan of the space borne GW detector, e.g., $2$-years for LISA. Under such conditions, the non-local degeneracies among intrinsic EMRI parameters become strongly intertwined with the additional degeneracies introduced by the bumpy parameter in the Bumpy-AK template. To tackle this complexity, two critical developments are required: first, a hierarchical search strategy capable of efficiently navigating the high-dimensional and wide prior of EMRI parameter spaces must be developed; second, ensemble averaging statistics must be rigorously integrated into this hierarchical framework in a self-consistent manner. This integrated approach must simultaneously resolve the dual-layer degeneracy problem while overcoming the needle-in-a-haystack obstacle. The complexity and multifaceted nature of these subjects require sustained scholarly engagement within a global-fit frameworks, and their systematic examination will be prioritized in our subsequent research agenda.
	\section*{ACKNOWLEDGMENTS} The work is supported by the National Key R\&D Program of China (Grant No. 2021YFC2203002), and
	the National Natural Science Foundation of China (Grants
	No. 12173071, No. 12473075). The computations were performed on the high-performance computers of School of fundamental Physics and Mathematical Sciences, Hangzhou Institute for Advanced Study, UCAS. We express our gratitude to RunQiu Liu for providing invaluable inspirations for our research. We also extend our appreciation to the administrator, Shutong Liu, for
	assisting us in utilizing the cluster effectively. Furthermore, we would like to acknowledge the insightful discussions that we had with Xian Chen, Peng Xu, Qun-ying
	Xie, Pin Shen, Qian-yun Yun, Runming Yao, Xue-hao Zhang, Shao-dong Zhao, Yi-yang Guo.

	
\end{document}